\documentclass{aastex}
\usepackage{emulateapj5,apjfonts,graphicx,psfig}

\def\ros{{\sl ROSAT }}

\def\ein{{\sl Einstein }}

\def\chandra{{\sl Chandra }}

\def\ergsec{\hbox{erg s$^{-1}$ }}
\def\ergcm{\hbox{erg cm$^{-2}$ s$^{-1}$ }}

\def\Msun{$M_{\odot}$ }

\def\it{\sl}

\slugcomment{accepted August 18th 2006 for publication in The Astrophysical Main Journal}

\shorttitle{X-RAY VARIABILITY IN $\theta^2$ ORI A}
\shortauthors{SCHULZ et al.}

\begin{document}

\title{X-ray Variability in the Young Massive Triple $\theta^2$ Ori A}

\author{
Norbert S. Schulz\altaffilmark{1}, 
Paola Testa\altaffilmark{1},
David P. Huenemoerder\altaffilmark{1},
Kazunori Ishibashi\altaffilmark{1},
Claude R. Canizares\altaffilmark{1}
}

\altaffiltext{1} {Kavli Institute for Astrophysics and Space Research, Massachusetts Institute of
Technology}

\email{nss@space.mit.edu}

\begin{abstract}
Massive stars rarely show intrinsic X-ray variability. The only O-stars 
credited to be intrinsically variable are $\theta^1$ Ori C due to effects from magnetic confinement
of its wind, and $\theta^2$ Ori A suspected of similar activity. Early \chandra observations have shown 
that the most massive star system in the 
Orion Trapezium Cluster, $\theta^2$ Ori A, shows rapid variability on time scales of hours.
We determine X-ray fluxes from observations with \chandra and find that the star
shows very strong variability over the last 5 years. We observed
a second large outburst of the X-ray source in November 2004 with the high resolution transmission
grating spectrometer on-board \chandra and compare the emissivity and line properties
in states of low and high flux. In the low state X-ray emissivities indicate temperatures
well above 25 MK. In the high state we find an extended emissivity distribution with high emissivities
in the range from
3 MK to over 100 MK. The outburst event in stellar terms is one of the most powerful ever
observed  and the most energetic one in the ONC with a lower total energy limit of
1.5$\times10^{37}$ ergs. The line diagnostics show that under the assumption that the He-like
ions are photoexcited the line emitting regions in the low states are as close as within  1 -- 2 stellar
radii from the O-star's photosphere, whereas the hard states suggest a distance of 3 -- 5 stellar radii.
We discuss the results in the
context of stellar flares, magnetic confinement, and binary interactions. By matching the 
dates of all observations with the orbital phases of the spectroscopy binary
orbit we find that the two outbursts are very close to the periastron passage of the stars.
We argue that the high X-ray states are possibly the result of reconnection events from magnetic interactions
of the primary and secondary stars of the spectroscopic binary. Effects from wind collisions seem
unlikely for this system. 
The low state emissivity and R-ratios strengthen the predicament that the X-ray emission is enhanced 
by magnetic confinement of the primary wind. We also detect Fe fluorescence indicative of the existence
of substantial amounts of neutral Fe in the vicinity of the X-ray emission.
\end{abstract}

\section{Introduction\label{sec:intro}}

The number of massive stars in the Orion Nebula Cluster is small
compared to other young star clusters of its class, but the few it harbors gained
much attention in recent years. At the forefront are the members of the 
Orion Trapezium, specifically the peculiar O5.5V star $\theta^1$ Ori C~\citep{schulz2001,gagne2005}, 
which have been under some scrutiny lately for their peculiar X-ray behavior
~\citep{schulz2003,stelzer2005}. Most of the massive stars
in the ONC show magnetic properties of some form, either intrinsically
or due to an unseen T Tauri companion.
With respect to intrinsic magnetic peculiarity there are so far three massive OB-stars with
direct evidence of a sizable magnetic field, $\theta^1$ Ori C, HD 191612, and
$\tau$ Sco ~\citep{donati2002, donati2006a, donati2006b}.
Other massive Trapezium stars exhibit similar X-ray spectral properties to $\theta^1$ Ori C, though
in none of these stars has a magnetic field been directly measured nor has any type
of X-ray variability been reported until recently~\citep{feigelson2002, stelzer2005}.
Possible magnetic properties of some sort of fossil origin have been suggested for most of them~\citep{schulz2001,
schulz2003}.

The primary star in $\theta^2$ Ori A is the second most massive early type star in the ONC after $\theta^1$ Ori C.
It is part of a massive system that
had its share of mystery in the early days of X-ray astronomy, when some studies 
proposed to identify the star as the optical counterpart of the X-ray source
2U 0526-06~\citep{barbon1972, aikman1974} and later of 4U 0531-05~\citep{bernacca1979}. One
of the main properties that made it a good candidate was its high mass function.
Its massive primary, a fifth magnitude O9.5V star~\citep{abt1991} of 
25~\Msun \citep{preibisch1999}, indicated the existence of
one or more massive companions. Specifically the lack of spectral signatures from a companion star
at the time hinted the possibility of the existence of a compact object.  
Today it is recognized that the system is unlikely to harbor a compact star but is considered
to be at least a massive triple system containing a visual companion 
with a mass of $< 8$~\Msun separated by $\sim 0.4"$~\citep{preibisch1999},
and a spectroscopic companion to the primary with 
a mass of likely $< 7 - 9$~\Msun \citep{abt1991}. The stellar types of both companion stars
are still highly uncertain and range from a massive early A-type star for the visual companion
to a B-type star as early as about B5 for the spectroscopic companion. Notable is also
the physical closeness of these massive companions, which amounts for the visual companion
to be 174 AU and the spectroscopic companion to be 0.47 AU~\citep{preibisch1999}, 
approximately 10 stellar radii off the primary. The system is likely as young as $\theta^1$ Ori C, i.e.
$\sim$ 0.3 Myr, which
is also reflected in the fact that the orbital motion of the spectroscopic binary 
and the O-star's rotational period are not yet synchronized. It cannot be much younger as it
already arrived on the zero-age main sequence, but also not much older through its association
with the Orion Trapezium.

In a recent study of X-ray emitting young stars in the Orion Nebula, ~\cite{feigelson2002} found that
$\theta^2$ Ori A and several early B-stars show short time scale variability. Especially $\theta^2$ Ori A exhibited
the most dramatic X-ray variability with a 50$\%$ flux drop over a time period of 
10 hours superimposed by multiple small flares with a few hour durations. Until today this
kind of variability in a massive system stands out as the most dramatic ever observed. 
The Chandra Orion Ultradeep Project (COUP) observations~\citep{getman2005} 
a few years later covered the X-ray intensity of that
star in several episodes for over 13 days and found it only weakly variable with the largest
variation around a factor 1.8~\citep{stelzer2005}. The observed variability in $\theta^2$ Ori A
clearly represents problems not only with the standard model of wind shocks in massive stars
but also with the proposed model of magnetically confined winds in these stars~\citep{gagne2005}.
\citet{feigelson2002} acknowledge that in some occasions large excursions in flux can be 
created by occasional large shocks as proposed by~\cite{feldmeier1997} but the temperature
profile during these events as observed is not consistent with such predictions.
In the case of $\theta^2$ Ori A a scenario was envisioned where magnetic reconnection events near the stellar
surface occur, which tentatively and qualitatively could explain the X-ray
emission observed in the O9.7Ib supergiant $\zeta$ Ori~\citep{waldron2001}.   

In this paper we present recent observations of $\theta^2$ Ori A observed with the high
energy transmission grating spectrometer (HETGS) onboard \chandra in a low and high flux state.

\section{Observations and Data Reduction\label{sec:obs}}

$\theta^2$ Ori A was observed with the \chandra HETGS (Canizares et al. 2005) on several occasions
between 1999 and 2005. Table 1 summarizes important observing
parameters for all observations with \chandra. Fourteen observations amount to a total exposure
of 1.2 Msec. All these observations provide at least a CCD quality X-ray spectrum
with spectral resolving powers between 6 and 50. Six observations were performed using the HETGS to
obtain highly resolved X-ray spectra with resolving powers between 180 and 1200. 
The observations performed during 1999 were also recorded at a different focal plane
temperature, which before February 2000 was -110 C and -120 C afterwards. This 
has some impact on the data reduction itself as all data run through the same 
reduction pipeline but with less developed calibration products for the -110 C case. 
Some of the now standardly used corrections for CCD charge
transfer inefficiency and gain would not be effective. 
The detected X-ray source in all of these observations was
never further away from the aimpoint than 3.0 arcmin, and less then 2.2 arcmin
for the grating observations. This is well within the margin where the
grating line spread functions do not suffer from a detectable degradation due 
to off-axis source position.

\begin{table*}[t]
\begin{center}
{\sc TABLE 1. All \chandra OBSERVATIONS THAT INCLUDE $\theta^2$ Ori A.} \\
\begin{tabular}{lcccccc}
            & & & & & & \\
\tableline
 Obsid &  Start Date & Start Time & Exposure & Instrument & HETG 1st order rate    & Phase\\
       & [UT]        & [UT]       & [ks]     &            & $10^{-2}$ cts s$^{-1}$ & range\\
\tableline
            & & & & & & \\
 0018 & Oct 12 1999 & 10:19:12 &  45.3 & ACIS-I&  --  & 0.86 -- 0.89 \\
 0003 & Oct 31 1999 & 05:47:21 &  52.0 & HETGS & 1.48 & 0.76 -- 0.79 \\
 0004 & Nov 24 1999 & 12:37:54 &  33,8 & HETGS & 3.91 & 0.92 -- 0.99 \\
 1522 & Apr 01 2000 & 17:31:12 &  37.5 & ACIS-I&  --  & 0.08 -- 0.10 \\
 2567 & Dec 28 2001 & 12:25:56 &  46.4 & HETGS & 3.92 & 0.99 -- 1.01 \\
 2568 & Feb 19 2001 & 20:29:42 &  46.3 & HETGS & 3.23 & 0.53 -- 0.55 \\
 4395 & Jan 08 2003 & 20:58:19 & 100.0 & ACIS-I&  --  & 0.33 -- 0.38 \\
 3744 & Jan 10 2003 & 16:17:39 & 164.2 & ACIS-I&  --  & 0.42 -- 0.51 \\
 4373 & Jan 13 2003 & 07:34:44 & 171.5 & ACIS-I&  --  & 0.54 -- 0.64 \\
 4374 & Jan 16 2003 & 00:00:38 & 169.0 & ACIS-I&  --  & 0.67 -- 0.77 \\
 4396 & Jan 18 2003 & 14:34:49 & 164.6 & ACIS-I&  --  & 0.79 -- 0.88 \\
 3498 & Jan 21 2003 & 06:10:28 &  69.0 & ACIS-I&  --  & 0.93 -- 0.96 \\
 4473 & Nov 03 2004 & 01:48:04 &  49.8 & HETGS & 2.80 & 1.00 -- 1.03 \\
 4474 & Nov 23 2004 & 07:48:38 &  51.2 & HETGS &19.31 & 0.96 -- 0.99 \\
\tableline
\end{tabular}
\end{center}
\label{observations}
\end{table*}

All observations were reprocessed using CIAO3.2 with the most recent CIAO
CALDB products. We used standard wavelength redistribution matrix files (RMF) 
and generated (ARFs) using the provided aspect solutions 
\footnote{see \url{http://asc.harvard.edu/ciao/threads/}}.
Note that for 
the HETGS spectra the RMF is fairly independent of focal plane temperature, however
the order sorting has to be adjusted to the different pulse height distributions
of the CCD at -110 C. Here we simply followed the steps applied in the data reduction
for the Trapezium stars described in~\cite{schulz2000}. For all the HETGS observations we 
generated spectra and analysis products for the medium energy gratings (MEG) +1 and -1
orders, as well as for the high energy gratings (HEG) +1 and -1 orders. 
Table 1 shows that for all HETGS observations except
for OBSID 4474, the average count rate is similar and low. Therefore for most parts of the 
analysis below we added all +1 and -1 orders of the MEG for OBSIDs 0003, 0004, 2567, 2568,
and 4473. The same procedure was applied to the HEG spectra. Obsid 4474 offers much higher
count rates and we proceeded with separate HEG and MEG spectra.

Contamination of the spectra from source confusion is to a large part mitigated by
the order sorting of the CCDs. The source itself is off the center of the
Orion Trapezium Cluster and its dispersed tracks do not as much interfere with
cluster core stars as, for example, during the analysis of $\theta^1$ Ori C~\citep{schulz2000}. In all the 
six HETGS data sets we only encountered a handful of doubtful emission lines, which
all could be eliminated by the fact that they appeared only once in the high energy
gratings (HEG) and medium energy gratings (MEG) orders or were identified as a COUP source. 
The HETGS spectral analysis was performed with ISIS~\citep{houck2000}.

\smallskip
\vbox{
\includegraphics[angle=0,width=8.5cm]{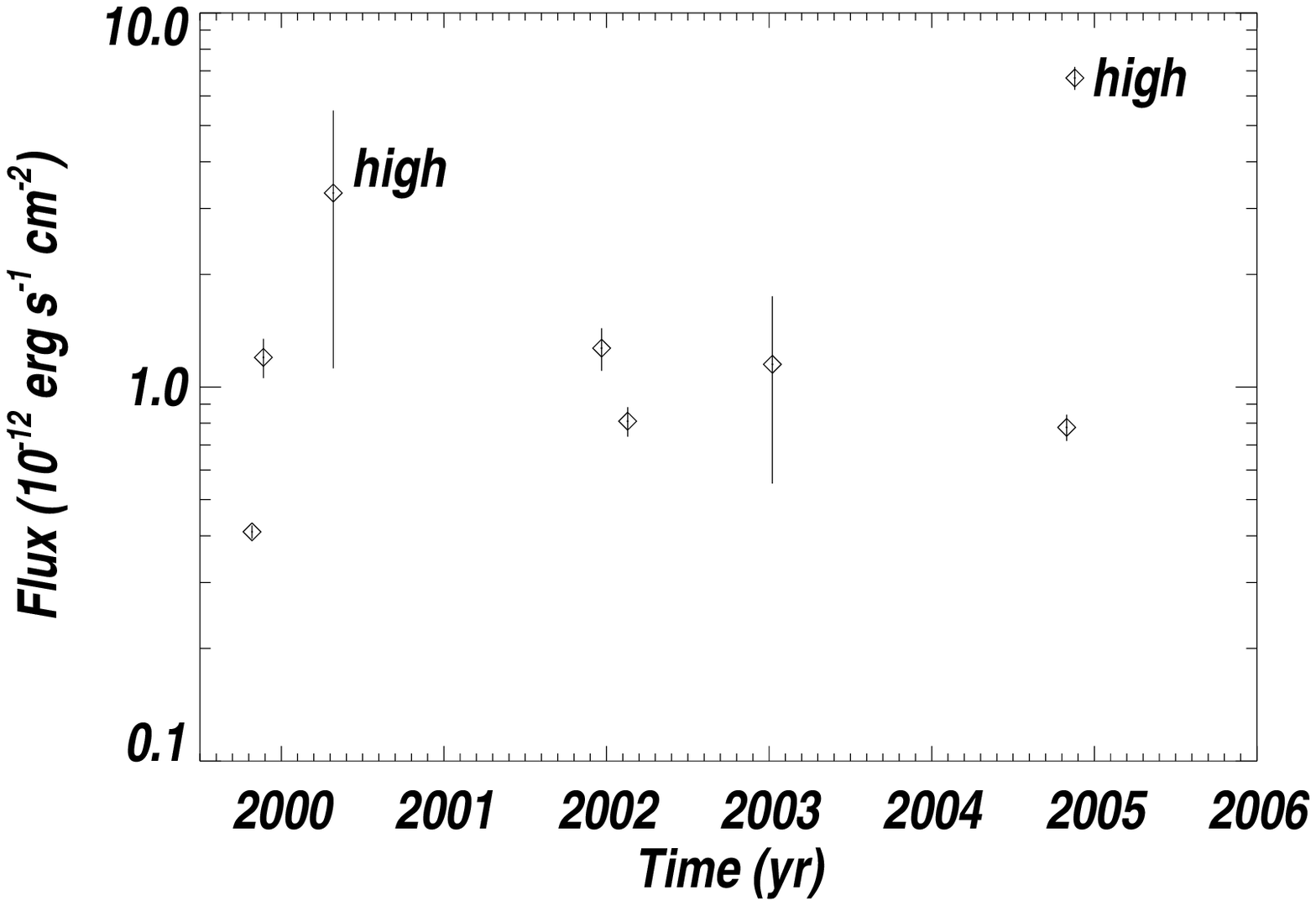}
\figcaption{This lightcurve was compiled from HETG 1st order fluxes where available. For
OBSID 1522 we used the estimates from~\citet{feigelson2002}, for the COUP data we used
the pileup model in ISIS~\citep{davis2001} to estimate the true count rate. 
The observations labeled with ``high'' also exhibit extreme
spectral hardening. Note that the value from the OBSID 1522 and COUP data has large systematic errors due
to the effects of large pileup.\label{longterm}}
}
\smallskip

\section{Long-term Variability\label{sec:ccd}}

Table 1 shows all \chandra observations of the resolved X-ray source. Unfortunately the 
source is bright enough to cause severe pileup in the observations performed without
the gratings. The piled rates range between 0.3 and 0.9 cts/frame, which is more than an order
of magnitude above the pile-up free regime. This problem was already
recognized by the previous studies, specifically for OBSID 1522 done in April 2000
~\citep{feigelson2002} and 
the COUP observations from January 2003~\citep{stelzer2005}. In order to mitigate the problem, these 
authors extracted a few percent of presumably unpiled flux from an annulus covering parts
the outer, low count wings of the point spread function. 
These results carry large uncertainties. We used the pileup model in ISIS~\citep{davis2001} and find that in the
case of OBSID 1522 the flux is slightly underestimated whereas in the COUP data the flux was
likely up to 40$\%$ lower.

\smallskip
\vbox{
\includegraphics[angle=0,width=8.5cm]{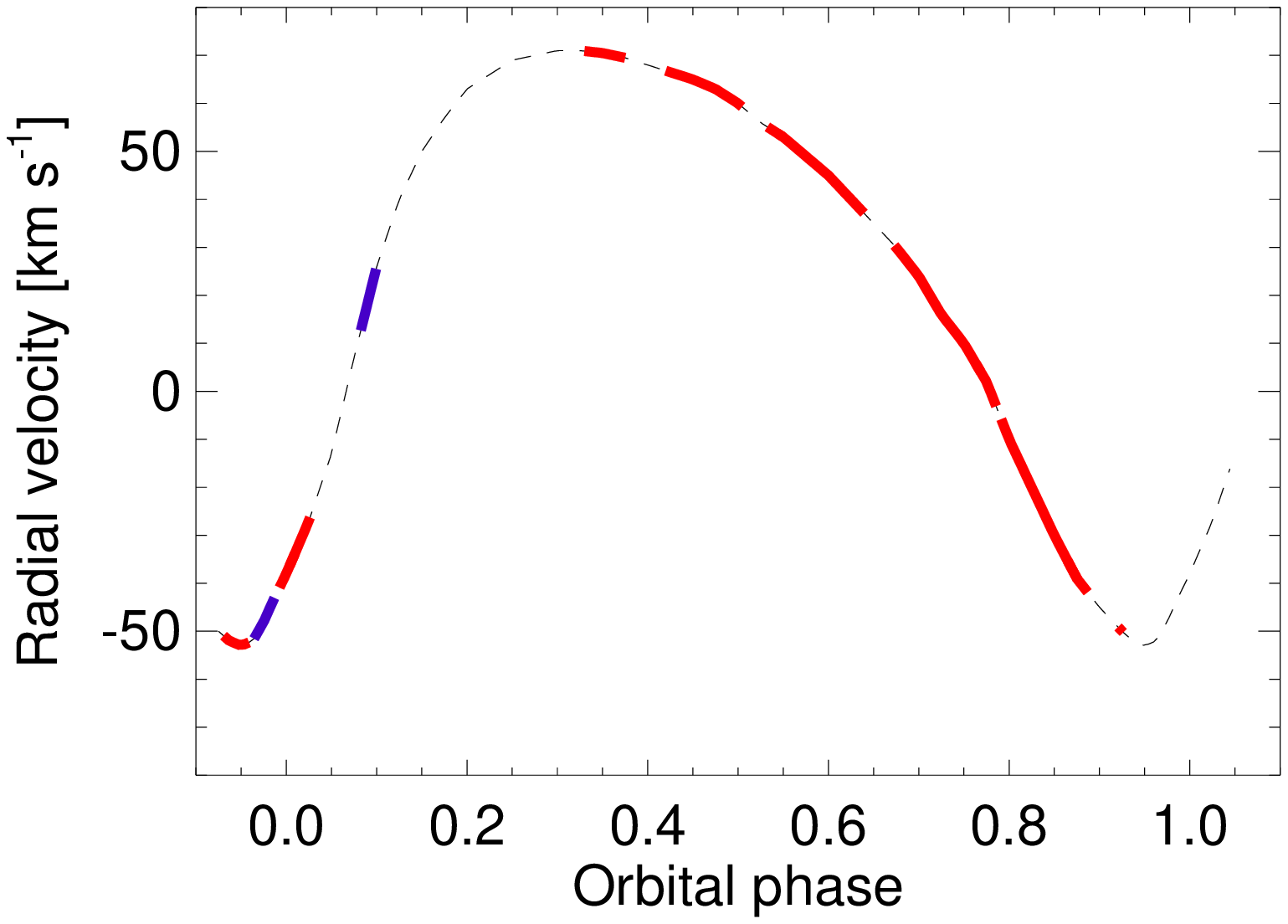}
\figcaption{Phase coverage of all \chandra observations with the radial velocity curve of the secondary
\citep{abt1991}. The red colored phase ranges correspond to the low state observations, the blue colors to the
high state observations.\label{radial}}
}
\smallskip

The observations
carried out with the HETGS do not have the pile-up problem in the dispersed spectra. Table 1 
lists the integrated count rates for all 1st order spectra. It is immediately recognized that 
there is a high level of variability in the source where the lowest and highest rate
differ by a factor 13. Specifically the last observation stands out with an extreme rate.
In an attempt to illustrate the long term behavior of the source we plotted
the HETGS 1st order fluxes together with the previously published estimates into a long-term
lightcurve (see Figure~\ref{longterm}). For the HETGS spectra we simply apply the averaged 
flux from all integrated low state grating spectra and scale it by the count rate from Table 1.
We do this because the statistics of a dispersed spectrum from single low state observations
is insufficient. For OBSID 1522 we used a time-averaged flux from ~\citet{feigelson2002}.
For the COUP data we also 
averaged the flux over the entire COUP exposure, even though the flux varied by almost
a factor two~\citep{stelzer2005}.

\smallskip
\vbox{
\includegraphics[angle=90,width=8.5cm]{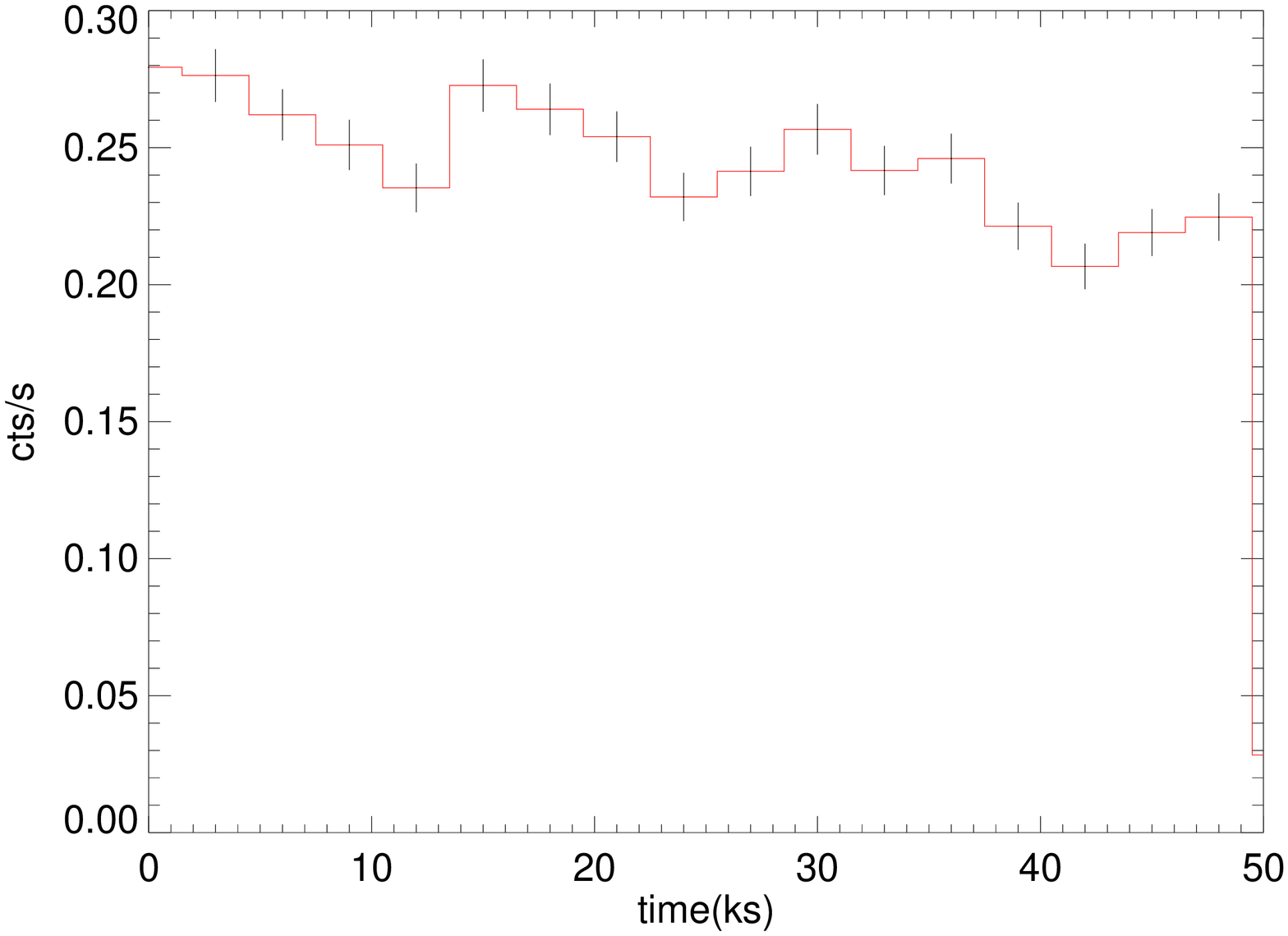}
\figcaption{The lightcurve of the outburst in Nov. 2004 (obsid 4474). The data are in 3 ksec
bins and from the summed HETGS 1st orders. It shows about a 30$\%$ decline in flux with intermittent
flaring.\label{newlight}}
}
\smallskip

In Figure~\ref{longterm} we identify two high flux and several lower flux states. The second
``high'' data point from 2004 clearly sticks out. The first data point marked
as ``high'' from 2000 does not appear as extreme. However its flare-like intrinsic lightcurve and its extreme
hardness, which is very comparable to the high 2004 observation, indicates that the flux here is 
underestimated. The hardness in the COUP data, on the other hand, is very close to 
the one observed in the HETG low flux observations. We therefore categorize the observation
in OBSIDs 1522 and 4474 as ``high state'' and the others as ``low state''.

\begin{figure*}[t]
\includegraphics[angle=0,width=16.5cm]{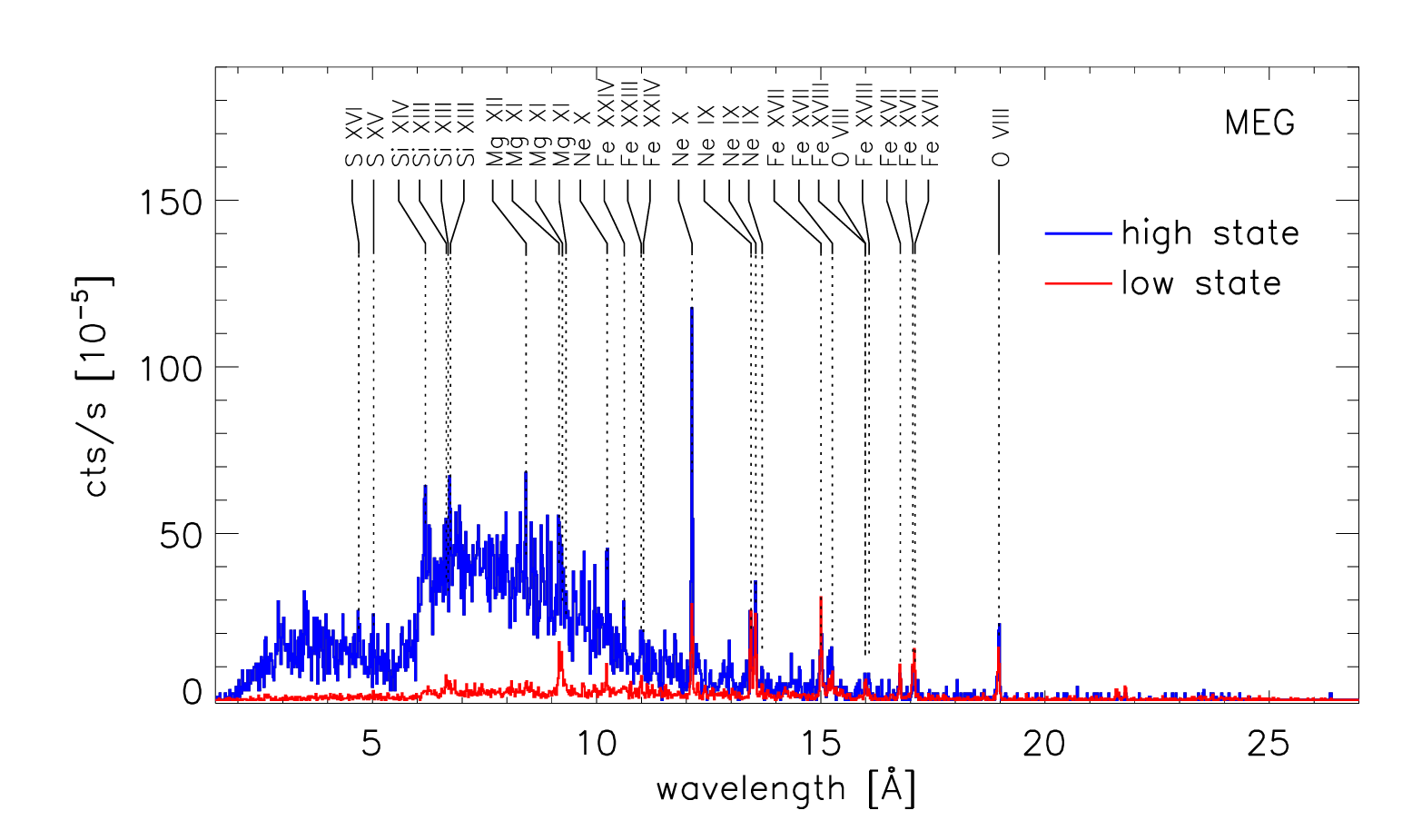}
\caption{The summed HETG spectra (MEG only) for the ``low state'' observations (red) and
OBSID 4474 for the ``high state'' (blue) with the identifications for the strongest lines.
The spectra are binned to MEG resolution (0.01~\AA).\label{hetgspectra}}
\end{figure*}

In order to further characterize the observed flux variability we go one step further and 
search for possible modulations of the observed X-ray with existing binary periods.
$\theta^2$ Ori A has a spectroscopic companion with a measured period of 20.974 days~\citep{abt1991}
and a projected distance 0.47 AU~\citep{preibisch1999}. This distance is an average one due
the determined orbital eccentricity of 0.33 and a large mass ratio of the two stars~\citep{abt1991}.
We recognized that the two high states apparently seem to overlap within a similar phase of
this orbital period. Figure~\ref{radial} shows the measured radial velocity curve
from ~\citet{abt1991}. It is highly asymmetric with a rapid change in speed near periastron passage.
companion. The periastron passage has an ephemeris of
HJD = 2440581.27~\citep{abt1991} from which we determined the orbital phase coverage of
all the X-ray observations performed with \chandra so far. 

The last column in Table 1 indicates the phase coverage for all the observations. Plotted 
in Figure~\ref{radial} the red segments indicate
observations which we associated with the X-ray low-state, the blue segments indicate high states.
Strikingly visible is a close coincidence of the high state emissivity with orbital
periastron passage. It appears that the match is not perfect, i.e. the two high states appear
right before and after phase 0 in a regime where the approach of the companion is closest, but low states
have been detected in between these high states.
Note that the lightcurve in~\citet{feigelson2002} indicates that the outburst started to wind
down after orbital phase 0.08, thus the onset likely happened around the phases 0.04 -- 0.05.
Figure~\ref{newlight} shows the lightcurve of the Nov. 2004 outburst during orbital phase 0.96 -- 0.99
where the X-ray flux stays nearly constant throughout the entire phase period. The flux appear to
be persistently high with a slow drop of about 30$\%$ from the start to the end. 
Future close tracking of the X-ray flux with orbital phase
of the spectroscopic binary is clearly necessary. 

\section{Analysis of HETG Spectra\label{sec:hetg}}

The analysis of the high resolution X-ray spectra was performed for low and high state
spectra. While we generally added positive and negative first orders for the MEG and HEG
separately, we also treated the five low state observations as one spectral state and 
added them together for statistical reasons. As can be seen in Table 1, the
total number of counts in the high state observation is still significantly higher than all low state observations
added together. The HETG spectra of the low state netted a total of 5807 counts and the 
high state 9007 counts. These translate in observed (absorbed) fluxes of 8.5$\pm 0.9 \times 10^{-13}$ \ergcm 
in the low state and 6.7$\pm 0.6 \times 10^{-12}$ \ergcm for the high state in the 2 -- 24~\AA~MEG band.
The spectra are shown in Figure~\ref{hetgspectra}.
Spectral diagnostics involves three parts. The first
part is devoted to the detection and determination of the emission line properties, i.e. line
identifications, line positions, and line fluxes. The second part uses the line fluxes as well
as underlying continuum to calculate corresponding emission measure distributions for the low and 
the high state. The third part then uses several emission lines and their ratios for to derive plasma
parameters.

\subsection{Emission Line Analysis\label{sec:emission}}

We use the Astrophysical Plasma Emission Database (APED)\footnote{http://hea-www.harvard.edu/APEC},
described by~\cite{smith2001} for the line identifications and line positions. We detected a total
of 39 lines in the high state and 36 lines in the low state (Table 2). The line
positions in all cases do not indicate significant shifts from the expected rest position. There is also
no notable difference between the line position in the two states. The line fluxes were measured
using custom software (PINTofAle~\citealt{kashyap2000}) and then corrected for the HETGS response. 
The Fe~XXVI, Fe~XXV,and Fe~K line measurements
are from HEG spectra, all the others from MEG spectra.

\begin{table*}[t]
\begin{center}
{\sc TABLE 2. LINE IDENTIFICATIONS AND FLUXES IN THE LOW AND HIGH STATES}
\begin{tabular}{lclccccc}
            & & & & & & &\\
\tableline
 Ion &  log T$_{peak}$ & $\lambda_0$ & $\lambda_{low}$ & $\lambda_{high}$ & Low State & High State \\
       & [K]  & [\AA]  & [\AA] & [\AA] & [$10^{-6}$ ph cm$^{-2}$ s$^{-1}$] & [$10^{-6}$ ph cm$^{-2}$ s$^{-1}$]\\
\tableline
            & & & & & & &\\
Fe~XXVI$^b$ & 8.1 & 1.780 (H)  &      --         & 1.766$\pm$0.008 &     --        &  9.0$^{+8.0}_{-2.2}$ \\
Fe~XXV$^b$  & 7.8 & 1.854 (L,H)&      --         & 1.856$\pm$0.005 &     --        &  9.3$^{+6.0}_{-2.8}$ \\
Fe~K        & --  & 1.937      &      --         & 1.934$\pm$0.004 &     --        &  7.3$^{+4.0}_{-2.3}$ \\
S~XVI$^b$   & 7.4 & 4.729 (H)  &      --         & 4.725$\pm$0.007 &     --        &  7.9$\pm$2.2 \\
S~XV$^b$    & 7.2 & 5.039 (L,H)& 5.029$\pm$0.008 & 5.030$\pm$0.004 & 0.87$\pm$0.58 & 12.2$\pm$3.0 \\
Si~XIV$^b$  & 7.2 & 6.182 (L,H)& 6.182$\pm$0.008 & 6.183$\pm$0.003 & 0.54$\pm$0.20 &  7.1$\pm$1.7 \\
Si~XIV$^a$  & 7.1 & 6.265      & 6.261$\pm$0.008 & 6.279$\pm$0.005 & 0.61$\pm$0.21 &  5.9$\pm$1.6 \\
Si~XIII$^b$ & 7.0 & 6.648 (LH) & 6.654$\pm$0.006 & 6.651$\pm$0.004 & 0.82$\pm$0.24 &  4.1$\pm$1.3 \\
Si~XIII     & 7.0 & 6.687      & 6.693$\pm$0.018 & 6.688$\pm$0.004 & 0.30$\pm$0.20 &  3.1$\pm$1.3 \\
Si~XIII     & 7.0 & 6.740      & 6.739$\pm$0.018 & 6.741$\pm$0.003 & 0.54$\pm$0.22 &  5.5$\pm$1.4 \\
Fe~XXIII    & 7.2 & 6.877      &      --         & 6.866$\pm$0.004 &     --        &  3.5$\pm$1.2 \\
Fe~XXIV     & 7.3 & 7.457      & 7.458$\pm$0.015 &       --        & 0.35$\pm$0.17 &      --      \\
Mg~XII$^b$  & 7.0 & 8.421 (L)  & 8.427$\pm$0.017 & 8.426$\pm$0.002 & 0.28$\pm$0.17 &  8.8$\pm$1.6 \\
Mg~XII$^a$  & 6.9 & 8.449      & 8.455$\pm$0.010 &       --        & 0.53$\pm$0.20 &      --      \\
Fe~XX$^a$   & 7.1 & 9.143      & 9.141$\pm$0.010 & 9.148$\pm$0.005 & 0.41$\pm$0.29 &  4.9$\pm$2.4 \\
Mg~XI$^b$   & 6.8 & 9.169 (H)  & 9.170$\pm$0.003 & 9.167$\pm$0.004 & 2.14$\pm$0.47 &  5.6$\pm$2.2 \\
Fe~XXI      & 7.1 & 9.194      & 9.208$\pm$0.005 & 9.193$\pm$0.004 & 1.21$\pm$0.44 &  4.5$\pm$2.2 \\
Mg~XI       & 6.8 & 9.231      & 9.239$\pm$0.003 & 9.235$\pm$0.004 & 2.28$\pm$0.49 &  5.4$\pm$1.7 \\
Fe~XX       & 7.0 & 9.282      & 9.287$\pm$0.010 & 9.269$\pm$0.005 & 0.83$\pm$0.39 &  4.6$\pm$1.9 \\
Mg~XI       & 6.8 & 9.314      & 9.315$\pm$0.030 & 9.315$\pm$0.007 & 0.57$\pm$0.35 &  3.9$\pm$1.9 \\
Fe~XX$^a$   & 7.1 & 9.500      &      --         & 9.499$\pm$0.005 &     --        &  4.7$\pm$1.2 \\
Ne~X        & 6.8 & 9.708      &      --         & 9.723$\pm$0.003 &     --        &  5.8$\pm$1.7 \\
Fe~XX$^a$   & 7.0 & 9.727      &      --         & 9.740$\pm$0.007 &     --        &  2.8$\pm$1.3 \\
Ne~X$^b$    & 6.8 & 10.239 (L,H)&10.230$\pm$0.004 &10.238$\pm$0.005 & 1.62$\pm$0.36 &  5.5$\pm$1.9 \\
Fe~XX$^a$   & 7.0 & 10.252      &      --         &10.252$\pm$0.004 &     --        &  5.9$\pm$1.9 \\
Fe~XXIV$^b$ & 7.3 & 10.619 (H)  &      --         &10.617$\pm$0.005 &     --        &  5.9$\pm$2.2 \\
Fe~XIX      & 6.9 & 10.645      &      --         &10.638$\pm$0.007 &     --        &  4.3$\pm$2.0 \\
Fe~XXIV$^b$ & 7.3 & 10.663 (H)  &      --         &10.660$\pm$0.007 &     --        &  2.3$\pm$1.7 \\
Fe~XXIII$^b$& 7.2 & 10.981 (H)  &      --         &10.993$\pm$0.007 &     --        &  5.6$\pm$2.4 \\
Fe~XXIV$^b$ & 7.3 & 11.029 (H)  &      --         &11.039$\pm$0.009 &     --        &  5.2$\pm$2.2 \\
Ne~X$^b$    & 6.8 & 12.134 (L,H)&12.129$\pm$0.002 &12.134$\pm$0.002 & 10.3$\pm$1.9  & 56.8$\pm$8.8 \\
Fe~XXIII    & 7.2 & 12.161      &12.148$\pm$0.004 &      --         &  3.8$\pm$1.3  &      --      \\
Fe~XIX$^b$  & 6.9 & 13.423 (L)  &13.425$\pm$0.002 &      --         &  0.57$\pm$0.48&      --      \\
Ne~IX$^b$   & 6.6 & 13.447 (L,H)&13.448$\pm$0.002 &13.450$\pm$0.005 & 13.6 $\pm$2.3 & 13.8$\pm$6.6 \\
Fe~XIX      & 6.9 & 13.462      &13.470$\pm$0.002 &      --         &  2.6 $\pm$1.5 &      --      \\
Fe~XIX$^b$  & 6.9 & 13.518 (L)  &13.523$\pm$0.002 &      --         &  4.4 $\pm$4.0 &      --      \\
Ne~IX       & 6.6 & 13.553      &13.556$\pm$0.002 &13.557$\pm$0.004 & 14.6 $\pm$2.0 & 22.8$\pm$7.6 \\
Fe~XIX      & 6.9 & 13.645      &13.647$\pm$0.002 &      --         &  1.3 $\pm$0.9 &      --      \\
Ne~IX       & 6.6 & 13.699      &13.702$\pm$0.010 &13.704$\pm$0.010 &  2.8 $\pm$1.2 &  5.8$\pm$4.9 \\
Fe~XVII$^b$ & 6.7 & 15.014 (L,H)&15.014$\pm$0.002 &15.020$\pm$0.004 & 30.2 $\pm$3.2 & 18.3$\pm$9.2 \\
Fe~XIX      & 6.9 & 15.198      &15.193$\pm$0.007 &15.210$\pm$0.008 &  6.6 $\pm$2.1 & 12.0$\pm$6.2 \\
Fe~XVII$^b$ & 6.7 & 15.261 (L,H)&15.264$\pm$0.004 &15.269$\pm$0.013 & 11.0 $\pm$2.6 & 12.1$\pm$8.3 \\
O~VIII      & 6.5 & 16.006      &16.008$\pm$0.006 &      --         &  6.5 $\pm$2.2 &      --      \\
Fe~XVIII$^b$& 6.8 & 16.071 (L)  &16.067$\pm$0.009 &      --         &  2.9 $\pm$1.8 &      --      \\
Fe~XVII$^b$ & 6.7 & 16.780 (L,H)&16.777$\pm$0.004 &16.768$\pm$0.008 & 12.3 $\pm$2.9 & 14.5$\pm$12.4\\
Fe~XVII$^b$ & 6.7 & 17.051 (L,H)&17.052$\pm$0.004 &17.050$\pm$0.010 & 12.6 $\pm$2.7 & 12.4$\pm$11.3\\
Fe~XVII$^b$ & 6.7 & 17.096 (L,H)&17.096$\pm$0.003 &17.100$\pm$0.015 & 17.7 $\pm$3.5 &  7.7$\pm$5.7 \\
O~VIII$^b$  & 6.5 & 18.969 (L,H)&18.972$\pm$0.004 &18.984$\pm$0.004 & 48.6 $\pm$8.3 & 71. $\pm$36. \\
O~VII   & 6.3 & 21.602 &21.605$\pm$0.008 &      --         & 15.6 $\pm$8.3 &      --      \\
O~VII   & 6.3 & 21.804 &21.801$\pm$0.005 &      --         & 23.  $\pm$13. &      --      \\
\tableline
\end{tabular}
\end{center}
\noindent
\footnotesize
a) These identification are either part of a blend or not unique due to temperature differences between
low and high state\hfill\break
b) Used for DEM solution: L (low state), H (high state); for the Fe XXV a lower flux limit was used in the
low state.
\normalsize
\label{lines}
\end{table*}

The lines are generally unresolved  and symmetric. In the regime of longer
wavelengths, i.e. for lines from Fe~XVII and Fe~XVIII ions as well as O~VIII
we observe some residual widths corresponding to about 300 km s$^{-1}$.
The region around the Mg~XI is heavily contaminated by various line blends from Fe and Ne ions.
Here we only list the strongest candidates from Fe. For a more complete discussion we refer
to~\cite{testa2004}.
Table 2 also includes a line detection of neutral Fe K fluorescence at 1.93~\AA\ during the high state. 
The line is detected close to the 2 $\sigma$ level. Its line position and width 
limit the fluorescence emission to very low ionization levels between
Fe~I and VII \citep{house1969}. 

\subsection{Emission Measure Analysis\label{sec:dem}}

The emission measure analysis was performed under the general
assumption of collisional ionization equilibrium (CIE) following
\citet{mazzotta1998} and abundances of
\citet{anders1989}.  A general outline for the computation of the
emission measure was given in \cite{huenemoerder2001} which has now
been successfully applied in several analyses of active coronal
sources~\citep{huenemoerder2003} and the hot star $\theta^1$ Ori C
\citep{schulz2003}. The volume emission measure (VEM) specifies the 
X-ray emissivity within a specified volume. The differential emission
measure (DEM) then is the VEM expressed as a function of temperature.
We constructed this DEM from line fluxes listed
in Table 2, spanning a range in log temperature of peak emissivity
from 6.6 to 8.1. The method derives the emission measure and
abundances simultaneously \citep{huenemoerder2003, huenemoerder2006}.
We fit the high and low states separately.  The range of uncertainties
due to the data quality are given by the width of the curves in the
figure, which was determined from Monte-Carlo iteration.
The lines used for the DEM calculation are marked in Table 2 for both states.
Uncertainties due to atomic data are not included.

Lines were excluded from the fit if they were expected to be sensitive
to density or photoexcitation (He-like triplet forbidden and
intercombination lines), or if they had large residuals (possibly due
to misidentification or blending).  The normalization was adjusted
post-facto in order to set the line-to-continuum ratio, since the DEM
solution was line-based.
The result for both states is shown in Figure~\ref{dems}. The red
curve is for the low state, and the blue curve the high state. The
thickness of the curves corresponds to the 68$\%$ confidence limits of
the DEM. The solid fill covers the ranges of line temperature peak
emissivities. 

The low state continuously declines in emissivity above $\log T
\mathrm{[K]} = 6.6$. The the high state shows strong peaks at 7.1 and
7.9.  We thus conclude, under the validity assumptions stated above,
that the quality of the emission between the two states changes and
the difference is not solely due to the volume of emission.

Abundances, though uncertain ( $\sim 20-40\% $), for the derived $DEM$
normalization were required to be about half Solar, except for Ne in
the low state which was near Solar.  There is weak evidence that Fe
and Ne were somewhat lower in the high state, with Fe changing from
0.65 to 0.15 and Ne from 1.0 to 0.5.  The results represent mean
values from the Monte-Carlo iteration.  The large range in the allowed emission measure
couples to a range in abundance (but with strong correlations between
different lines and temperatures).  We do not find the abundance
trends conclusive given the uncertainties present in the data and
modeling, but they are suggestive of compositional changes in the
plasma and warrant more careful examination. In $\theta^1$ Ori C 
abundances in the DEM suggested more nearly solar composition but also
showed a low Fe abundance~\citep{schulz2003,gagne2005}.

\subsection{Line Diagnostics\label{sec:ratios}}

Certain line ratios can be used to diagnose optical depth, densities, and 
sources of radiation. In the case of helium-like triplets it is the metastable 
forbidden line component that is sensitive to plasma density or, in the presence of
radiation fields, to photoexcitation \citep{kahn2001, blumenthal1972, gabriel1969}. 
We resolve the triplets in O~VII, Ne~IX, Mg~XI, Si~XIII, and S~XV, though in the 
case of S XV, the range of sensitivity is quite limited (see ~\citealt{porquet2001}
for an overview). We compute two ratios, an R-ratio which is the ratio between 
the forbidden and intercombination line flux, and the ratio of the H- to He-like resonance lines. 
While the former diagnoses plasma densities in the range between 10$^{10}$ cm$^{-3}$
and 10$^{15}$ cm$^{-3}$ or a UV field between 900 -- 1500~\AA, the
latter is sensitive to local plasma temperature. For the determination of these ratios
one has to be careful with the placement of the underlying continuum as well as
interfering line blends. In nearly all cases the continua are strong and well determined
(see Figure~\ref{hetgspectra}), except for O~VII in the high state spectrum, which
was too weak to be detected. Note that the temperature increase for the high state in the cool part of Figure~\ref{dems}  
is shifted towards higher temperatures. In the low state it remains unclear where the cool temperature
peak actually is. Interference by line blends is strong in 
Ne~IX and Mg~XI. For Ne~IX Table 2 shows several known L-shell lines from Fe ions,
which are well resolved and identified, but only for the low state. 
The Mg~XI band has been modeled in detail by
~\citet{ness2003}. The Mg~XI
triplet, on the other hand, has interfering lines in both states. These are not
identified and are likely satellite lines not included in APED.
It should also be noted that except for O~VII, the forbidden lines are detected in 
all investigated ions.

\smallskip
\vbox{
\clearpage
\includegraphics[scale=0.55]{f5.eps}
\figcaption{The emission measure distribution $[\mathrm{cm}^{-3}]$ for
  the low (red) and the high (blue) state as
  calculated from the emission line fluxes in Table 2. The upper and
  lower line boundaries are the 68$\%$ confidence limits.
\label{dems}}
}
\smallskip

The H- to He-like resonance line ratios and R-ratios from the He-like triplets are summarized in Table 3.
Interpretation of the R-ratio requires additional information about the radiative environment. 
A sensitivity to density can only be claimed if sources of photoexcitation are
ruled out or included in the calculation. On the other hand, if there are 
such sources, one may use the ratio to place limits to the distance 
from the radiation source to the line forming region as has been demonstrated
for the hot star $\zeta$ Ori \citep{waldron2001}. Under the assumption that 
the X-rays are produced in the O-star wind, we calculated the dependence
of the R-ratio on the distance from the stellar photosphere. For this we assumed 
a stellar surface temperature of 30000 K for the O9.5 star and used 
photoexcitation and decay rates from~\citet{blumenthal1972}. 
Figure~\ref{distance} then shows these functions for O~VII, Ne~IX, Mg~XI, and Si~XIII.
The values of the R-ratios are then plotted as diamonds onto these curves for
the low state (left) and the high state (right). The 1~$\sigma$ range
for the ratios are highlighted as thick lines. It can be seen that the ratios 
of O~VII and Si~XIII are not well constrained. For Si~XIII
the range spreads over the entire sensitivity range of distance
curve. In the case of O~VII, the R-ratio
is in a range of the curve that is not very sensitive to distance. 
The deduced values place the ions within $\sim$5 stellar radii in both 
states. Ne~IX and Mg~XI are very well constrained. Ne~IX shows no change between
high and low state. The range of radii for Mg~XI seems to lie closer to the stellar surface
in the low state. At these distances, the R-ratio is always strongly affected by
the O-stars radiation field and is not a useful diagnostic of plasma density.

\begin{table*}
\begin{center}
{\sc TABLE 3. G- AND R-RATIOS FOR THE LOW AND HIGH STATES}\\
\begin{tabular}{lccccccc}
            & & & & & &\\
\tableline
     &  \multicolumn{3}{c}{low state}  & & \multicolumn{3}{c}{high state} \\
Ion  & Ly$\alpha$/r  & T [MK]  & $f/i$ & & Ly$\alpha$/r & T [MK] & $f/i$ \\
\tableline
    &                   &       & &               & &      &\\
O~VII   &  3.1 $\pm$ 1.75   &  [2.2 -- 4]  &    --           & &  --             &  --      & --\\
Ne~IX  &  0.76 $\pm$ 0.19  &  [3.4 -- 4.3]  & 0.20 $\pm$ 0.09 & & 4.1 $\pm$ 2    &  [5,4 -- 10]   & 0.25 $\pm$ 0.20 \\
Mg~XI  &  0.13 $\pm$ 0.08  &  3.7 -- 5.2]  & 0.26 $\pm$ 0.15 & & 1.6 $\pm$ 0.7  &  [7 -- 16] & 0.82 $\pm$ 0.53 \\
Si~XIII  &  0.66 $\pm$ 0.29  &  [8 -- 13]  & 1.8 $\pm$ 1.4  & & 1.7 $\pm$ 0.8  &  [8 -- 18]  & 1.77 $\pm$ 0.83 \\
\tableline
\end{tabular}
\end{center}
\label{gratios}
\end{table*}

\smallskip
\includegraphics[angle=0,width=8.5cm]{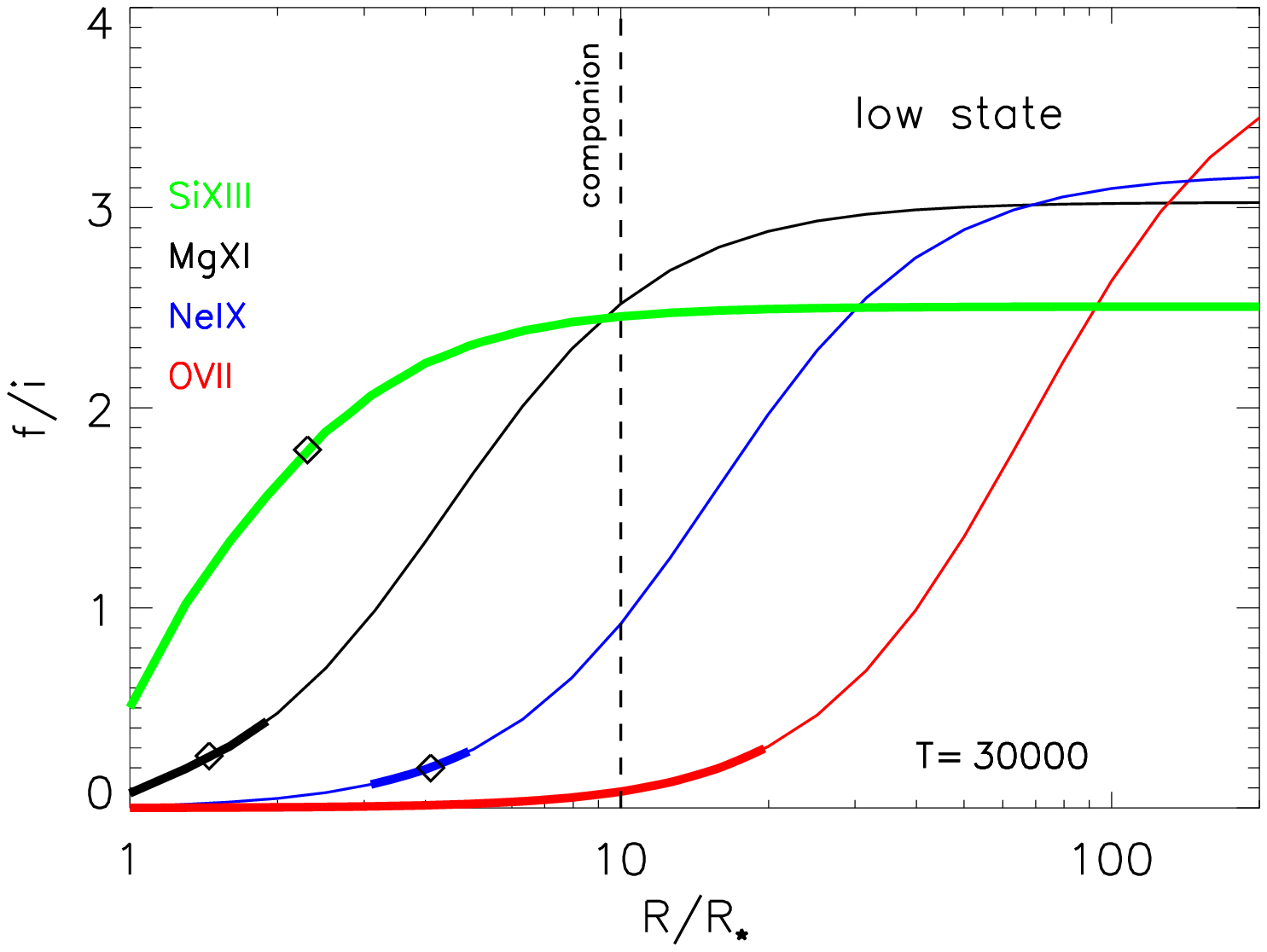}
\hspace{0.3cm}\includegraphics[angle=0,width=8.5cm]{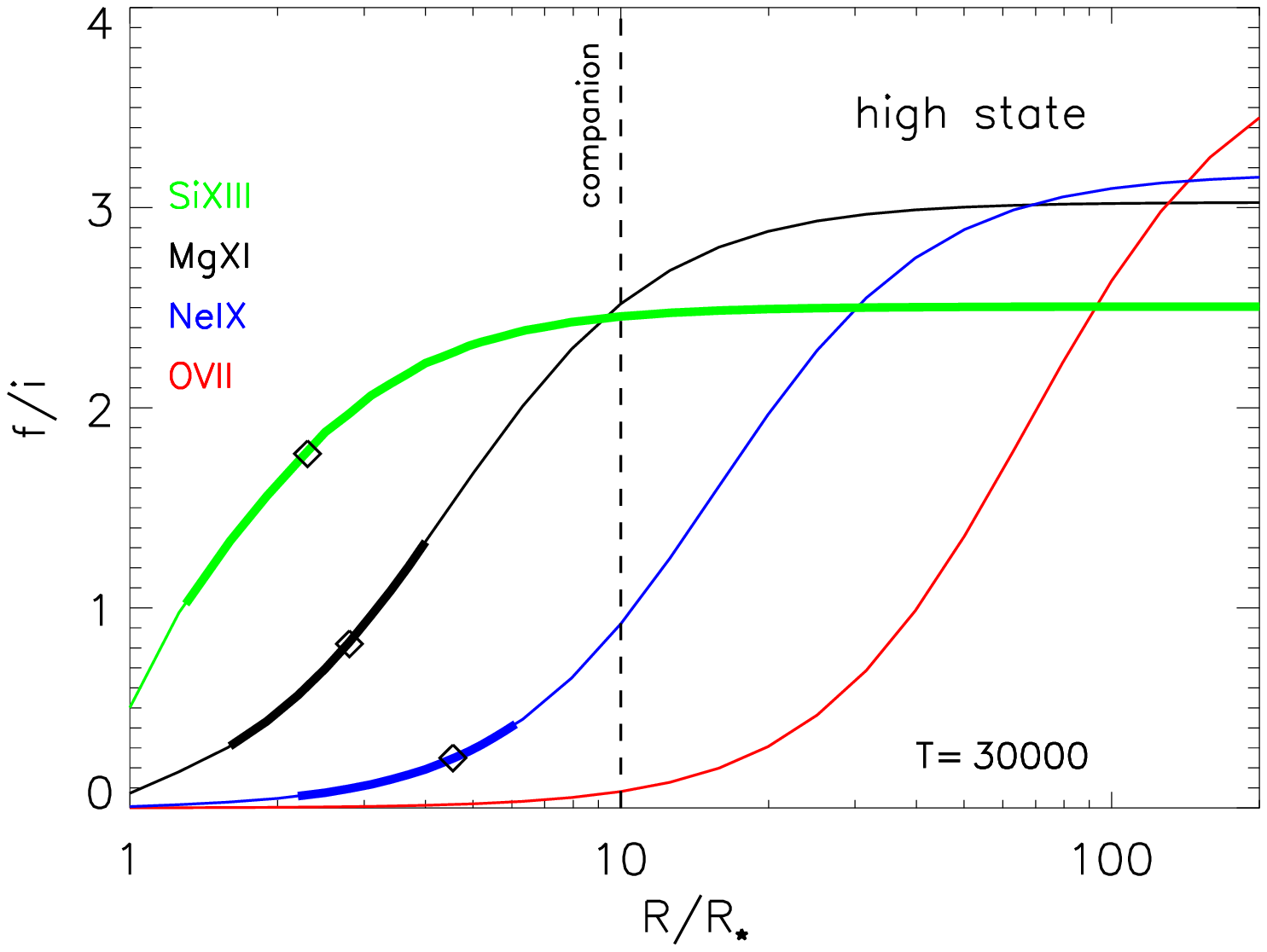}
\figcaption{The dependence of the R-ratios of the O~VII, Ne~IX, Mg~XI, and Si~XIII
triplets on the distance to the stellar surface of the primary component in
$\theta^2$ Ori A for the low (top) and the high (bottom) state. The highlighted line portions
represent the projected 1 $\sigma$ confidence limits of the measured ratios. In the
case of O~VII there is only an upper limit in the low state. The surface temperature
of the O9.5 star was assumed to be 30000 K. The vertical hatched
line marks the distance of the spectroscopic companion as listed in~\cite{preibisch1999}.
\label{distance}}
\smallskip

\section{Discussion\label{sec:disc}}

This study reveals several new aspects and details about the X-ray emission of
young massive stars. $\theta^2$ Ori A proves
to be a new study case for variable emission from massive systems. The analysis confirms the
findings by \citet{feigelson2002} that this system undergoes strong excursions in X-ray flux.
The outburst reported on in this study appears much more dramatic in terms of
flux levels and spectral hardening even if one accounts for pileup which heavily distorted the data
in the 2000 event. Before we discuss the possible nature of the variable X-ray emission we summarize the
results of our analysis. 

Over a period of about 5 years the X-ray source showed variability in flux of over an order of magnitude.
During the last outburst the range of observed plasma temperatures changed from 3 -- 30 MK to a range
of 3 -- $>$100 MK. The X-ray luminosities changed from 2.6$\times10^{31}$ \ergsec to 1.7$\times10^{32}$ 
\ergsec which corresponds to a change of the $L_x/L_{bol}$ ratio from $\sim 10^{-7}$ to 
$\sim 10^{-6}$. Under the assumption that all O-stars are X-ray emitters of some form and the one in 
$\theta^2$ Ori A is no exception, we have to assume
that the radiation field of the primary O9.5V star is the main source of destruction of the 
forbidden lines in the He-like triplets. The dominance of the UV field puts the X-ray source
close, i.e. within a few stellar radii of the primary's photosphere.  
As a consequence a possible contribution
from high density to the destruction of the lines is limited to the order
of $ 10^{10}$ cm$^{-3}$ for the low and high state at the most. High densities may only be feasible,
if the X-ray source is at a very large distance from the O-star and the entire emissivity is not
associated with the O-star wind. We are not discussing this possibility as consequently the 
O-star itself would not be an X-ray emitter which based on the observational record of X-rays from
early type stars is unlikely. 
The observed emissivities in these two states are then understood not only in terms of
a change in the energy balance due to a dramatic change in temperatures, but also
due to a vastly increased emissivity volume in the high state. 
Last but not least, the coincidence
that the two outbursts appear close to the nearest approach of the spectroscopy companion
suggests that they are related to some form of interaction.

In the following we discuss these results under various aspects. First we 
put the outbursts into the context of variability observed
in the ONC so far with \chandra. We then attempt to qualitatively interpret the 
observed X-ray emissivity
with respect to intrinsic physical environments such as magnetic confinement of winds,
magnetic reconnection, and binary interactions.
Last we discuss possible origins of the Fe K fluorescence during the high state.

\subsection{Hard X-ray Variability and Reconnection\label{sec:hard}}

\smallskip
\vbox{
\includegraphics[angle=0,width=8.5cm]{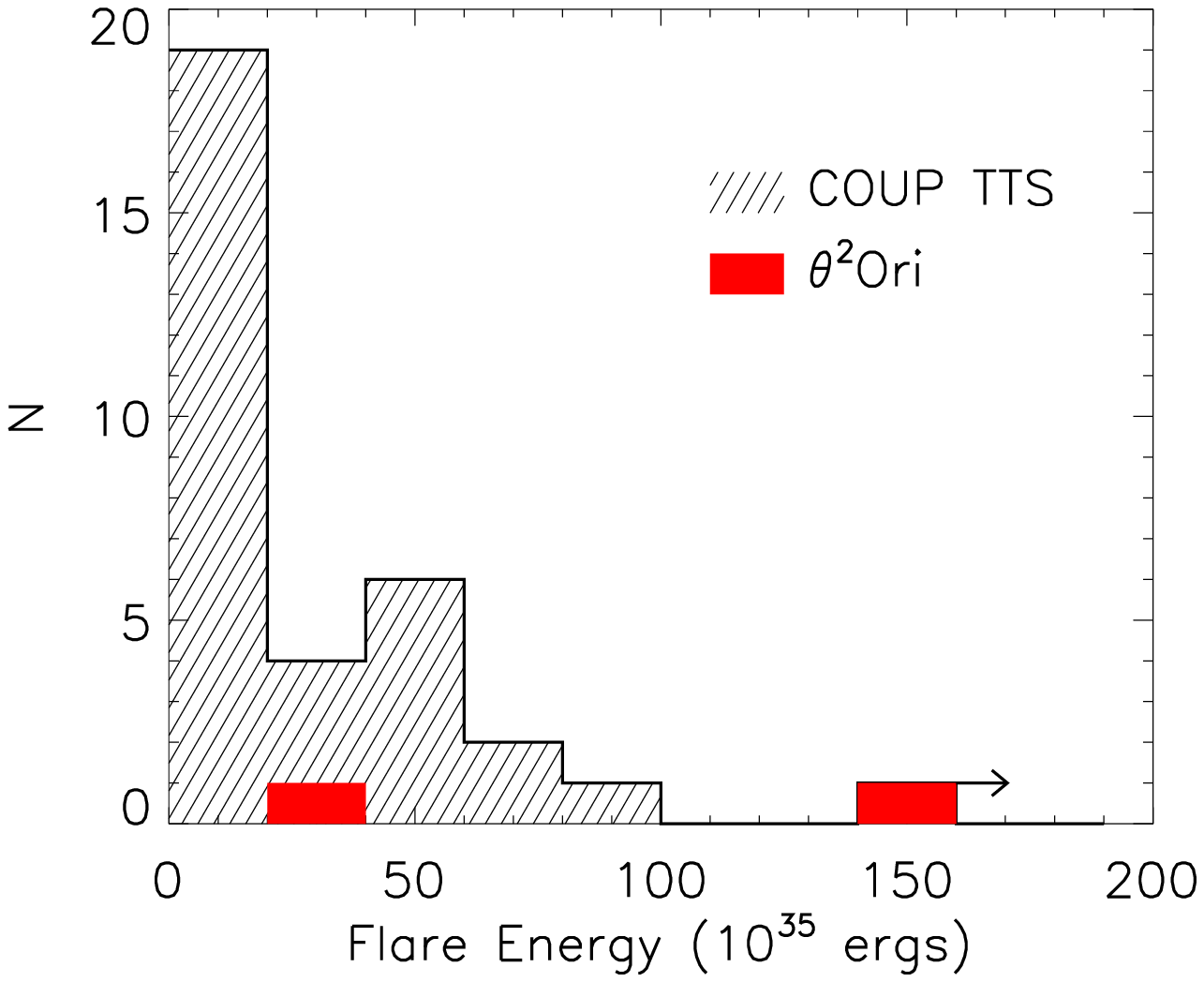}
\figcaption{The number distribution of the integrated energy of all bright flares
in the COUP study~\citep{favata2005}.
\label{flare}}
}
\smallskip

Figure~\ref{flare} illustrates the extraordinary nature of the observed variability. We integrated
the total energy of all bright flares of young late type stars 
analyzed in \citet{favata2005} in the COUP data and 
sorted them by flare energy. We integrated the flares from their observed maximum using the observed decay time.
All the COUP flares have well established peak luminosities,
temperatures, and cooling time constants. The statistics then shows that most
flares are found near a total energy of less than 10$^{36}$ ergs, with the brightest one barely reaching
10$^{37}$ ergs. 
In this respect the outburst in $\theta^2$ Ori A from the year 2000 does not look particularly
impressive with a total energy of 3$\times10^{36}$ ergs. However, the 
lightcurve in \citet{feigelson2002} does not   
appear like the flares seen in late type stars  which show a short rise and exponential decay.
In both outbursts, though, the lightcurves showed their highest
fluxes at the beginning of each observation with some form of decline afterwards.
Thus we assumed the flux at the beginning of the 
lightcurve as the starting point and used the rest of the lightcurves to estimate its decay time. 
The integrated energy should then be considered a lower limit because we do not really know the actual 
peak occurrence. The second 
outburst in 2004 in this respect clearly distinguishes itself from all the other observations as well
as generic flares (see Figure~\ref{newlight}). The emissivity remains high over the 
entire exposure with some decline in flux. Under the assumption that
the late 2004 event was some form of flare event, the resulting lower limit to the total energy 
is 1.5$\times10^{37}$ ergs.

Flares this powerful are unprecedented in normal stellar systems and it is very likely that
these high states are of a different origin. A somewhat less powerful flare has been recently uncovered
in \ros data of the massive binary $\sigma$ Ori E~\citep{groote2004}. 
For a flare event of coronal type the conditions have to be rather extreme indeed.
Dissipation of large electrical
currents associated with reconnection events realigning the field leads to the
release of magnetic energy. Having established that the hard X-ray source during the outburst is likely located 
at a few stellar radii from the O-star's surface, a corresponding magnetic loop structure
seems extraordinarily large. Even though the star is expected to rotate 
relatively fast with rotation period between 2.8 days~\citep{preibisch1999} and
3.5 days~\citet{kaper1998} the density of the wind at a few stellar radii is likely too low
to cause the substantial field dragging necessary to explain the observed energy release. 
Such rotation rates produce surface velocities of about 1500 km s$^{-1}$ and
we should have seen some line broadening from additional shocks or turbulence if the 
wind would be affected.
Applying scaling laws of stellar flares from the Sun also show that reconnection
events heat the ambient plasma on time scales of
several minutes. Cooling times should be of the order of only a few hours. The outburst of 2004 
persists for almost an entire day suggesting that there is some continuous heating. 

\subsection{Effects of Magnetic Confinement\label{sec:magnetic}}

It is generally assumed that X-ray emission from hot stars 
originates from shocks in line driven winds.  
O-stars of all spectral and luminosity classes have so far been observed in X-rays. 
In hot stars like $\theta^1$ Ori C, $\tau$ Sco, and likely $\zeta$ Ori energy is thought to be deposited into the
wind through magnetic channeling resulting in hard X-ray emission. Since
the main component of $\theta^2$ Ori A is of O~9.5 V type, it seems warranted to
associate at least some of the observed X-ray emission likely with the wind of the 
O-star. It has also been recognized in recent years that temperatures much in excess of 
15 MK cannot be the result of standard wind shocks but rather the result of more complex
structures such as magnetic confinement or interacting winds. 

\vbox{
\smallskip
\includegraphics[angle=0,width=9.5cm]{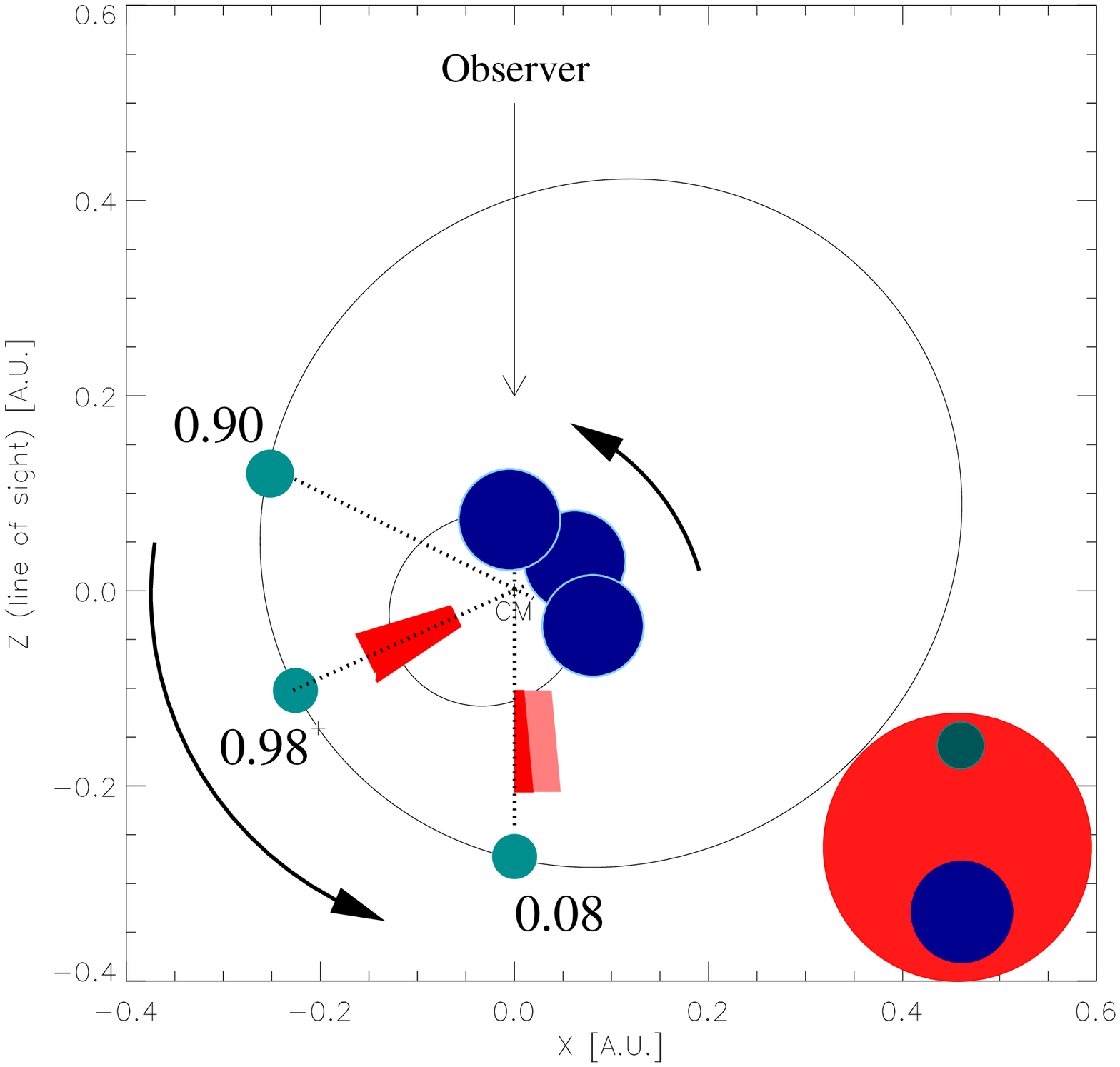}
\figcaption{Schematic illustration of configurations of the binary using the orbital 
parameters given by~\cite{abt1991}. The
system is plotted face-on, the inclination towards the observer is estimated to be 65$^{\circ}$.
The black arrows indicate the direction of motion around the center of mass. The sizes of the
stellar companions are scaled about proper to the orbit dimensions. The blue circle represents the O9.5V star,
the green circle the spectroscopic companion. The configuration at orbital phase 0.90 marks the projected
start of outburst activity, the one at 0.98 marks the outburst observed in November 2005, the one at
0.08 marks the start of the outburst reported by~\citet{feigelson2002} and the projected end of activity. 
We used the start of the
flux decline in the lightcurve of the 2000 outburst to set the end point. Symmetrically to periastron
we then projected the start point of the activity. 
The filled red areas mark the anticipated locations of the hard outburst X-ray emission. The lighter red shade
indicates the decline of the 2000 outburst. The interaction region is large compared to the size of the primary
star and its moving through the line of sight should not affect the view of the X-ray source by much.   
This is visualized in the cartoon in lower right corner, which is 
now the line of sight view of the observer at 0.08 orbital phase.
\label{orbit}}
}
\smallskip

In comparison with other hot stars (see Figure 5 in \citealt{wojdowski2005}) 
the DEMs of $\theta^2$ Ori A show several novel aspects. Both, the high and 
low states are different in their emissivity from what is expected from shocks in an unobstructed
line driven stellar wind. In the low state (red curve) the X-ray spectrum and its
corresponding DEM compare well with X-ray emission from $\tau$ Sco with the 
difference that we see not only strong emissivity at log T[K] = 7, but also emissivity 
at much lower temperatures. 
Unobstructed winds seem rather to peak below log T[K] = 6.5 
with a steady decline of emissivity above about 3 -- 5 MK. Two more facts strengthen
the comparison with $\theta^1$ Ori C and $\tau$ Sco: the emission
lines are similarly narrow and the origin of the emission
is likely more than one stellar radius above the photosphere of the star. 
In both cases magnetic fields have been detected, most recently for $\tau$ Sco~\citep{donati2006b}.
The emission from $\theta^2$ Ori A in the low state then should be discussed in the 
context of some form of magnetic channeling of the O-star wind.  

In the high state the DEM of Figure~\ref{dems} differs from anything we have observed before, i.e. a 
high emissivity from log T[K] = 6.5 to 8.0 with a tail towards even higher temperatures.
Peak emissivities beyond log T[K] = 7.5 are known from flares
in stars with active coronae and from $\theta^1$ Ori C. The X-ray  emissivity level in the latter
star is an order of magnitude larger than observed at any level 
in $\theta^2$ Ori A owing to a stronger wind and a high magnetic field of the order of $\sim$1 kG~\citep{gagne2005}. 
In $\theta^1$ Ori C the soft emissivity is weak 
and the bulk of the emissivity is growing towards higher temperatures. In $\theta^2$ Ori A the outburst
DEM rises high at lower temperatures and then remains high over a large range of high temperatures.  

Since a magnetic field has not yet been directly measured so far in $\theta^2$ Ori A 
we rely on projections from the models presented by \citet{babel1997} and \citet{gagne2005}
for the cases of IQ Aur and $\theta^1$ Ori C. Clearly, the contrast between these
two cases could not be larger, i.e. the mass loss rates of a few times $10^{-10}$ \Msun yr$^{-1}$ and
$10^{-7}$ \Msun yr$^{-1}$ differ by three orders of
magnitude and the surface temperatures (13000 K and 42000 K, respectively) 
by over a factor three. $\theta^2$ Ori A's values
are in somewhere between. Mass loss rates of less than $10^{-8}$ \Msun yr$^{-1}$
and a stellar surface temperature of 30,000 K suggest a
strong wind. Under assumption that
the wind has a moderate 
velocity, i.e. v$_{\infty}$ of the order of 1000 km s$^{-1}$, we would obtain a confinement
parameter ($\eta = B^2R^2/\dot Mv_{\infty}$) of 10 for the 
existence of a magnetic field of $>$ 0.6 kG. 

Quantitative modeling by \citet{uddoula2002} suggests that a confinement parameter of 10
or above should result in effective confinement. 
The model also predicts that most X-rays are produced in a relatively small region where the
density of the hot gas is high, beyond $10^{12}$ cm$^{-3}$. To reach such high densities
and temperatures, confined regions form within one ot two Alfv\'en radii from the stellar surface where 
field lines are still mostly closed and perpendicular to the magnetic equatorial plane.
Thus confinement places the X-ray source close to the stellar surface of the primary. 
For $\theta^2$ Ori A the R-ratios in
Figure~\ref{distance} put the distance of the X-ray source less than 1 stellar radius in
the case of Mg~XI and  2 -- 4 stellar radii in the case of Ne~IX
above the stellar surface.
The confinement model and data in the case of $\theta^1$ Ori C place the region of emission
within $\sim$2 stellar radii~\citep{gagne2005}.

Our data
seem to rule out densities that are as high as $10^{12}$ cm$^{-3}$. However the low state spectrum
indicates much lower emissivity at high temperatures than observed in $\theta^1$ Ori C
suggesting that such high densities may not be required.
The R-ratios are not entirely conclusive with respect to the location of the emitting region  
above the stellar surface and predictions from magnetic confinement. 
In the low state, the Mg~XI R-ratio
seems to indicate such a closeness, the Ne~IX ratio on the other hand could allow for the larger range.
In the high state the emission region is more like 3 - 5 R$_{star}$ from the O-star surface. Confinement in 
the latter case seems difficult to achieve. Thus a simple confinement scenario as proposed for 
$\theta^1$ Ori C and likely $\tau$ Sco is feasible for the low state but is problematic for the 
high state.

\subsection{Binary Interactions\label{binary}}

\subsubsection{Colliding Winds\label{colliding}}

We also have to consider the fact that all stars in the $\theta^2$ Ori A system are of early
type and thus all are sources of radiatively driven winds. The proximity of the spectroscopic companion
could then give rise to some form of wind interaction.
Colliding wind scenarios to produce X-ray emission~\citep{prilutskii1976, stevens1992} 
have been proposed since the launch of
\ein where a survey of WR stars revealed that WR-O binary system are systematically
brighter in X-rays~\citep{pollock1987}. Recent observations of O-star binaries such as HD93403
(O5.5I + O7V~\citealt{rauw2002}), HD 159176~\citep{debecker2004}, and HD152248
(O7III+O7.5III~\citealt{sana2004}) involve O-stars of very early type or luminosity classes III
and above or both. 
Other cases involving O dwarfs and B-stars such as HD 206267 and $\iota$ Ori have so far not
been conclusive in X-rays~\citep{pittard2000, schulz2003a}. Colliding wind scenarios are usually
associated with exceptionally strong winds with
mass loss rates exceeding $10^{-6}$ \Msun yr$^{-1}$ and terminal velocities of
$>$ 2000 km s$^{-1}$. Unusually bright X-rays from very young O-star binaries
have been observed in young star-forming regions containing many massive stars and in which
strong stellar winds create a diffuse interstellar gas~\citep{townsley2003}.

The case of $\theta^2$ Ori A is quite different. Its O9.5V star may produce a mass loss of the 
order of only about 1$\times10^{-8}$ \Msun yr$^{-1}$. 
Its companion has likely a much lower rate. Recent research also suggests that mass loss rates
in Galactic O-type Stars could be even smaller than generally assumed~\citep{fullerton2006}.
A colliding wind scenario is expected to differ with respect to systems like WR 140 or HD152248 where
mass loss rates are over 1000 times higher. In 
HD152248 the separation of the two stars is always close, within 2 primary stellar radii. 
Figure~\ref{orbit} shows that even at closest approach, the secondary in $\theta^2$ Ori A is still over
6 primary radii apart from
the O-star surface and wind densities should not exceed a few times $10^{9}$ cm$^{-3}$. However,
wind models also suggest that the density cannot be much lower either. At much lower densities
heating is provided by collisionless shocks like in WR 140~\citep{pollock2005} but would require
unrealistically high emission volumes on scales in excess of 6 AU. In the optimistic case
of effective wind compression which may allow for densities up to $10^{10}$ cm$^{-3}$ the volumes would be
on the scale of only a few 0.01 AU. But then we still need an average collision speed  $>> 2000$ km s$^{-1}$
and near 100$\%$ collision efficiency, again extremely unrealistic conditions.

Furthermore, in colliding wind systems luminosity is inversely proportional to the distance
from the companion to the shock interface, which at an orbital eccentricity of the order of 
0.33 as suggested~\citep{preibisch1999} would predict a slow and gradual increase and decrease
over significant fractions of the binary orbit. 
The fact that we have low state observations intermediate to the observed high states points to more
explosive origins. We therefore rule out scenarios that involve interacting winds only.

\subsubsection{Interacting Magnetic Fields\label{fields}}

Sect. \ref{sec:hard} already demonstrated that 
simple magnetic reconnection scenarios related to an O-star's magnetic field alone seem unfeasible.
On the other hand, the proximity of the
outbursts to the periastron passage suggests the possibility of two body magnetic interaction events.
Figure~\ref{orbit} schematically illustrates the anticipated orbital geometry. In the two body case,
the X-ray source is located in between the two stars, which near periastron are separated
by about 6 -- 7 primary radii.
With an integrated emissivity during the outburst of 2.1$\times10^{55}$ cm$^{-3}$
the emitting volume has a size of (0.2 AU)${^3}$ assuming realistic plasma densities. Magnetic
reconnection thus has to happen on quite large scales. We can estimate an upper
limit to the magnetic energy release by calculating the heat produced by the drag of
the primary field lines due to the approaching companion to

\begin{equation}
\frac{d E_{mag}}{dt} = \eta \Omega_{rad} B_{surf}^2 R_d^3,
\end{equation}

\noindent
where $B_{surf}$ is the field strength of the O-star primary, $\Omega_{rad}$ is the absolute
difference between the angular velocity
of the incoming secondary and the primary rotation, $R_d^3$ the size of the interaction region. The parameter 
$\eta$ is 
an efficiency factor acounting for reconnection efficiency for specific geometries and 
accounting for the fact that not all magnetic field lines participate in a specific release event.
A similar approach has been applied by
~\cite{shu1997} to estimate the energy release in star-disk interactions in T Tauri stars. 
For a primary field of $\sim$ 0.6 kG and an
incoming speed of the companion of 250 km s$^{-1}$ we obtain a maximum energy release of
$\sim 10^{36}$ \ergsec. This value should be compared to the observed luminosity during the
outburst, which is a few times $10^{32}$ \ergsec. This indicates that $\eta$ can be very small.
It can also mean that our basic assumptions are incorrect. However,
given the large emissivity volume and a 
likely weak companion field this may not be unrealistic. Also given the many unknowns in this
system with respect to such a complex scenario, we cannot draw more detailed conclusions. 
More data and modeling clearly
seem warranted.

\subsection{Origins of the Fe Fluorescence\label{sec:fluor}}

In the high state we also detect a few photons which we attribute to Fe K fluorescence at 1.93~\AA. 
The outburst produces a substantial amount of X-ray continuum flux above
7.11 keV (1.74~\AA), the value of the photoelectric absorption edge. Fluorescence
emission emerges from cool, mostly neutral, and optically thick materials 
located near ionizing environments. 
In the X-ray band fluorescence is most efficient for Fe atoms because the high atomic number
mitigates autoionization through Auger electrons. Fe fluorescence
has therefore the highest yields and strong Fe K lines are observed in X-ray binaries with large accretion disks
\citep{gottwald1995},
high column density material in winds~\citep{sako1999, schulz2002}, 
proto- and PMS stellar disks~\citep{tsujimoto2005}, but also in the Sun
\citep{bai1979, parmar1984}. 

The fact that we observe fluorescence in such a system is a novelty, if one disregards
high mass X-ray binaries, which also contain stars of early type with strong winds but also
neutron stars or stellar black holes. Here
strong winds from evolved supergiants are almost completely ionized by a 
compact X-ray source giving rise to fluorescence
emission from cool, optically thick clumps dispersed throughout the wind
\citep{sako1999,schulz2002,boroson2003}. \citet{feldmeier1995}
pointed out that winds can have distinct regions of clumpy old materials specifically in the outer regions
of the wind. Recent work designed to explain the low attenuation of X-ray lines quite recently
led to studies involving a general porosity of stellar winds~\citep{owocki2006}. 
Although the wind 
in the O9.5 dwarf may have such clumpiness on some scale, we do not consider these
as a possible source for the fluorescence because they do not provide a sufficient mass with respect
to the incident luminosity. The X-ray source in Vela X-1 is of the order of 10$^4$ times brighter
than the outburst in $\theta^2$ Ori A.

We investigated a variety of possible sources and conclude that none of the sites alone can account
for the observed photons. Young systems contain a lot of debris and circumstellar material
even after several million years after formation and given the projected age of the system
of less then 1 Myr one should expect plenty of circumstellar matter in the system. However, the fact that
the O-star entered the main sequence also means that its powerful wind has evaporated most of 
circumstellar material within the spectroscopic binary with a probability of some leftover disk with
the visual companion. This leaves us with the optically thick photospheres of the companion as well as 
any possible disk structure that evolved from the wind confinement. 

In order to estimate the expectation of fluorescence from the O-star's photosphere we 
derived an efficiency for fluorescence from the stellar surface \citep{bai1979}
as $\epsilon \sim 2*I_{line}/I_{c}$, where $I_c$ is the incident hard continuum flux
above 7.11 keV and $I_{line}$ the measured fluorescence line flux. The model in \citet{bai1979}
was designed to estimate photospheric fluorescence from the Sun, but since its geometry only
scales with the stellar radius, we may relate the efficiency to the behavior in the Sun. 
For $\theta^2$ Ori A we measured an efficiency $\epsilon = 0.064 \pm 0.045$. 
For the Sun $\epsilon$ = 0.04 is expected at the 
surface and drops steeply above one solar radi~\citep{parmar1984}. Fluorescence from the photosphere alone is thus
likely not sufficient to explain the observed line flux which is consistent with the non-detection 
Fe fluorescence in $\theta^1$ Ori C~\citep{schulz2003} where the hard X-ray flux was even higher. This
shows that the fluorescence has to include other sources mentioned above.

\section{Conclusions\label{conclusions}}

The X-ray emission from 
$\theta^2$ Ori A showed unusually strong variability over the last 5 years. Its last outburst
in Nov. 2004 was energetically one of the most powerful recorded so far in a stellar system, consisting
of only early type stars. The outbursts appear to be near periastron passage indicating
a connection between a close approach of the companion and outburst activity. 

The primary star is of O9.5V type and 
by its nature associated with a sizable radiatively driven wind. The fact that we observe narrow 
X-ray lines with maybe some turbulent broadening of $\sim$ 300 km s$^{-1}$, a hard emissivity distribution,
and R-ratios that indicate that the X-ray source is a close as $<$ 2 primary stellar radii to the O-star's
surface lets us conclude that the X-ray emission in the {\it low} state is enhanced by magnetic confinement
of the wind possibly similar to the model which was successfully applied to $\theta^1$ Ori C~\citep{gagne2005}.
We do not observe evidence for a hybrid behavior as in $\theta^1$ Ori C~\citep{schulz2003}.

The X-ray lines in the {\it high} state appear mostly unresolved, again with some evidence of turbulent broadening
in the soft lines. However, the emissivity distribution is now greatly enhanced at temperatures above 50 MK
and the R-ratios indicate that the bulk of the X-ray emission arises from larger distance from the stellar surface.
Based on our favored conclusion of a magnetic origin of the low state emission and the association of the outbursts with
a close approach of the massive spectroscopic binary star we propose a scenario of intermittent, strong and 
persistent reconnection events on likely various time scales as a consequence of interacting magnetic fields
of the primary and secondary star. Effects
from wind collisions are not favored mainly for reasons related the the lack of sufficient wind density
and the explosive appearance of the activity. 
 
We also observe fluorescence emission for which we cannot assess a sufficiently potent candidate source.
We therefore conclude that we observe a combination of many origins, such as the photospheres of the stars,
confined wind disks, and some remnants from protostellar activity.

Support for this work was provided by the National
Aeronautics and Space Administration through the
Smithsonian Astrophysical Observatory contract SV3-73016 to
MIT for Support of the Chandra X-Ray Center, which is
operated by the Smithsonian Astrophysical Observatory for
and on behalf of the National Aeronautics Space
Administration under contract NAS8-03060.


\begin{thebibliography}{8}

\bibitem[\protect\astroncite{{Abt} et~al.}{1991}]{abt1991}
{Abt}, H.~A., {Wang}, R., {Cardona}, O. 1991, \apj, 367, 155

\bibitem[\protect\astroncite{{Aikman} \& {Goldberg}}{1974}]{aikman1974}
{Aikman}, G.~C.~L. and {Goldberg}, B.~A. 1974, \jrasc, 68, 205

\bibitem[\protect\astroncite{{Anders} \& {Grevesse}}{1989}]{anders1989}
{Anders}, E. \& {Grevesse}, N. 1989, \gca, 53, 197

\bibitem[\protect\astroncite{{Babel} \& {Montmerle}}{1997}]{babel1997}
{Babel}, J., {Montmerle}, T. 1997, \aap, 323, 121

\bibitem[\protect\astroncite{{Bai}}{1979}]{bai1979}
{Bai}, T. 1979, \solphys, 62, 113

\bibitem[\protect\astroncite{{Barbon} et~al.}{1972}]{barbon1972}
{Barbon}, R., {Bernacca}, P.~L. \& {Tarenghi}, M. 1972, \nat, 240, 182

\bibitem[\protect\astroncite{{Bernacca} \& {Bianchi}}{1979}]{bernacca1979}
{Bernacca}, P.~L., {Bianchi}, L. 1979, \aap, 75, 61

\bibitem[\protect\astroncite{{Boroson} et~al.}{2003}]{boroson2003}
{Boroson}, B., {Vrtilek}, S.~D., {Kallman}, T. \& {Corcoran}, M. 2003,\apj, 592, 516

\bibitem[\protect\astroncite{{Blumenthal} et~al.}{1972}]{blumenthal1972}
{Blumenthal}, G.~R., {George}, R., {Drake}, G.~W.~F., {Tucker}, W.~H. 1972, \apj, 172, 205

\bibitem[\protect\astroncite{{Canizares} et~al.}{2005}]{canizares2005}
{Canizares}, C.~R., {Davis}, J.~E., {Dewey}, D., {Flanagan}, K.~A., {Galton}, E.~B., {Huenemoerder}, D.~P.,
{Ishibashi}, K., {Markert}, T.~H., {Marshall}, H.~L., {McGuirk}, M., {Schattenburg}, M.~L., {Schulz}, N.~S.,
{Smith}, H.~I., \& {Wise}, M. 2005, \pasp, 117, 1144 

\bibitem[\protect\astroncite{{Cassinelli} et~al.}{1989}]{cassinelli1989}
{Cassinelli}, J.~P., {Schulte-Ladbeck}, R.~E., {Abott}, M., \& {Poe}, C.~H. 1989,
Proceedings of the IAU Coll. No. 113, Kluwer, Dortrecht, p. 121

\bibitem[\protect\astroncite{{Cassinelli} et~al.}{1994}]{casinelli1994}
{Cassinelli}, J.~P., {Cohen}, D.~H., {MacFarlane}, J.~J., {Sanders}, W.~T.,
{Welsh}, B.~Y. 1994, \apj, 421, 705

\bibitem[\protect\astroncite{{Cassinelli} \& {Miller}}{1998}]{cassinelli1998}
{Cassinelli}, J.~P., {Miller}, N.~A. 1998, Proceedings of the IAU Coll. No. 169,
Springer, Berlin, Heidelberg, New York, p. 169

\bibitem[\protect\astroncite{{Cassinelli} et~al.}{2001}]{cassinelli2001}
{Cassinelli}, J.~P., {Miller}, N.~A., {Waldron}, W.~L., {MacFarlane}, J.~J., \&
  {Cohen}, D.~H. 2001, \apjl, 554, L55

\bibitem[\protect\astroncite{{Davis}}{2001}]{davis2001}
{Davis}, J.~E. 2001, \apj, 562, 575

\bibitem[\protect\astroncite{{De Becker} et~al.}{2004}]{debecker2004}
{De Becker}, M., {Rauw}, G., {Pittard}, J. M., {Antokhin}, I. I., {Stevens},
I. R., {Gosset}, E., {Owocki}, S. P. 2004, \aap, 416, 221

\bibitem[\protect\astroncite{{Donati} et~al.}{2002}]{donati2002}
{Donati}, J.-F., {Babel}, J., {Harries}, T.~J., {Howarth}, I.~D., {Petit}, P., \& {Semel}, M. 2002, \mnras, 333, 55

\bibitem[\protect\astroncite{{Donati} et~al.}{2006a}]{donati2006a}
{Donati}, J.-F., {Howarth}, I.~D., {Bouret}, J.-C., {Petit}, P., {Catala}, C., {Landstreet}, J.
2006, \mnras, 365, 6 

\bibitem[\protect\astroncite{{Donati} et~al.}{2006b}]{donati2006b}
{Donati}, J.-F., {Howarth}, I.~D., {Jardine}, M.~M. et al., \mnras, submitted

\bibitem[\protect\astroncite{{Favata} et~al.}{2005}]{favata2005}
{Favata}, F., {Flaccomio}, E., {Reale}, F., {Micela}, G., {Sciortino}, S., {Shang}, H., {Stassun}, K.~G., \&
{Feigelson}, E.~D. 2005, \apj Suppl., 160, 469

\bibitem[\protect\astroncite{{Feigelson} et~al.}{2002}]{feigelson2002}
{Feigelson}, E~.D., {Broos}, P., {Gaffney}, J.~A. III, {Garmire}, G., {Hillenbrand}, L.~A., 
{Pravdo}, S.~H., {Townsley}, L., \& {Tsuboi}, Y. 2002, \apj, 574, 258

\bibitem[\protect\astroncite{{Feldmeier} et~al.}{1997}]{feldmeier1997}
{Feldmeier}, A., {Puls}, J., {Pauldrach}, A.~W. 1997, aap, 322, 878

\bibitem[\protect\astroncite{{Feldmeier}}{1995}]{feldmeier1995}
{Feldmeier}, A. 1995, \aap, 299, 523

\bibitem[\protect\astroncite{{Friend} \& {Abbott}}{1986}]{friend1986}
{Friend}, D.~B., {Abott}, M. 1986, \apj, 311, 701

\bibitem[\protect\astroncite{{Fullerton} et~al.}{2006}]{fullerton2006}
{Fullerton}, A.~W., {Massa}, D.~L. \& {Prinja}, R.~K. 2006, \apj, 637, 1025

\bibitem[\protect\astroncite{{Gabriel} \& {Jordan}}{1969}]{gabriel1969}
{Gabriel}, A.~H., {Jordan}, C. 1969, \mnras, 145, 241

\bibitem[\protect\astroncite{{Gagn\'e} et al.}{2005}]{gagne2005}
{Gagn\'e}, M., {Oksala}, M.~E., {Cohen}, D.~H., {Tonnesen}, S.~K.,
{ud-Doula}, A., {Owocki}, S.~P., {Townsend}, R.~H.~D., \& {MacFarlane}, J.~J. 2005, \apj, 628, 986

\bibitem[\protect\astroncite{{Gagn\'e} et al.}{2005}]{gagne2005b}
{Gagn\'e}, M., {Oksala}, M.~E., {Cohen}, D.~H., {Tonnesen}, S.~K.,
{ud-Doula}, A., {Owocki}, S.~P., {Townsend}, R.~H.~D., \& {MacFarlane}, J.~J. 2005, \apj, 634,712 

\bibitem[\protect\astroncite{{Garmire} et~al.}{2000}]{garmire2000}
{Garmire}, G., {Feigelson}, E.~D., {Broos}, P., {Hillenbrand}, L., 
{Pravdo}, S.~H., {Townsley}, L., \& {Tsuboi}, Y. 2000, \aj, 120, 1426

\bibitem[\protect\astroncite{{Getman} et~al.}{2005}]{getman2005}
{Getman}, K.~V., {Flaccomio}, E., {Broos}, P.~S., {Grosso}, N., {Tsujimoto}, M., {Townsley}, L., {Garmire}, G.,
{Kastner}, J.~H., {Li}, J., {Harnden}, F.~R., {Wolk}, S., {Murray}, S., {Lada}, C.~J., {Muench}, A., {McCaughrean}, M.~J.,
{Meeus}, G., {Daminai}, F., {Micela}, G., {Sciortino}, S., {Bally}, J., {Hillenbrand}, L.~A., {Herbst}, W., 
{Preibisch}, T., \& {Feigelson}, E.~D. 2005, \apj Suppl., 160, 319

\bibitem[\protect\astroncite{{Gottwald} et al.}{1995}]{gottwald1995}
{Gottwald}, M., {Parmar}, A.~N., {Reynolds}, A.~P., {White}, N.~E., {Peacock}, A.,  {Taylor}, B.~G. 1995,
\aaps, 109, 9

\bibitem[\protect\astroncite{{Groote} \& {Schmitt}}{2004}]{groote2004}
{Groote}, D. \& {Schmitt}, J.~H.~M.~M. 2004, \aap, 418, 235

\bibitem[\protect\astroncite{{Hartmann} \& {MacGregor}}{1982}]{hartmann1982}
{Hartmann}, L., {MacGregor}, K.~B. 1982, \apj, 282, 591

\bibitem[\protect\astroncite{{Houck} \& {Denicola}}{2000}]{houck2000}
{Houck}, J.~C. \& {Denicola}, L.~A. 2000, 
ASP Conf. Ser. 216: Astronomical Data Analysis Software and Systems IX, p. 591

\bibitem[\protect\astroncite{{House}}{1969}]{house1969}
{House}, L.~L. 1969, \apj Suppl., 18, 21

\bibitem[\protect\astroncite{{Howk} et al.}{2000}]{howk00}
{Howk}, J.~C., {Cassinelli}, J.~P., Bjorkman, J.~E., Lamers H.~J.~G.~L.~M. 2000, \apj, 534, 348

\bibitem[\protect\astroncite{{Huenemoerder} et al.}{2001}]{huenemoerder2001}
{Huenemoerder}, D., {Canizares}, C.~R., \& {Schulz} N.~S. 2001, \apj, 559, 1135

\bibitem[\protect\astroncite{{Huenemoerder} et al.}{2003}]{huenemoerder2003}
{Huenemoerder}, D.~P., {Canizares}, C.~R., {Drake}, J.~J., \& {Sanz-Forcada}, J. 2003, \apj, 595, 1131

\bibitem[\protect\astroncite{{Huenemoerder} et al.}{2006}]{huenemoerder2006}
{Huenemoerder}, D.~P., {Testa}, P., \& {Buzasi}, D. 2006, \apj, 650, in press 

\bibitem[\protect\astroncite{{Kahn} et~al.}{2001}]{kahn2001}
{Kahn}, S.~M., {Leutenegger}, M.~A., {Cottam}, J., {Rauw}, G., {Vreux}, J.-M.,
  {den Boggende}, A.~J.~F., {Mewe}, R., \& {G{\" u}del}, M. 2001, \aap, 365,
  L312

\bibitem[\protect\astroncite{{Kaper}}{1998}]{kaper1998}
{Kaper}, L. 1998, Proceedings of the IAU Coll. No. 169,
Springer, Berlin, Heidelberg, New York, p. 193

\bibitem[\protect\astroncite{{Kashyap} \& {Drake}}{2000}]{kashyap2000}
{Kashyap}, V. \& {Drake}, J.~J. (2000), B.A.S.I., 28, 475

\bibitem[\protect\astroncite{{Kastner} et~al.}{2002}]{kastner2002}
{Kastner}, J.~H., {Huenemoerder}, D., {Schulz} N.~S., {Canizares}, C.~R., \&
{Weintraub}, D.~A. 2002, \apj, 567, 434

\bibitem[\protect\astroncite{{Liedahl}}{1998}]{liedahl1998}
{Liedahl}, D.~A. 1998,in Lecture Notes in Phys. 520, X-ray Spectroscopy in 
Astrophysics, ed. J. van Paradijs \& J.~A.~M. Bleeker (Berlin, Springer), pp. 189

\bibitem[\protect\astroncite{{Lucy}}{1982}]{lucy1982}
{Lucy}, L.~B. (1982), \apj, 255, 286

\bibitem[\protect\astroncite{{Lucy} \& {White}}{1980}]{lucy1980}
{Lucy}, L.~B., {White}, R.L. 1980, \apj, 241, 300

\bibitem[\protect\astroncite{{Mazzotta} et~al.}{1998}]{mazzotta1998}
{Mazzotta}, P., {Mazzitelli}, G., {Colafrancesco}, S. \& {Vittorio}, N. 1998, \aaps, 133, 403 

\bibitem[\protect\astroncite{{Ness} et~al.}{2003}]{ness2003}
{Ness}, J.-U., {Brickhouse}, N.~S., {Drake}, J.~J. \& {Huenemoerder}, D.~P. 2003, \apj, 598, 1277

\bibitem[\protect\astroncite{{Owocki} et~al.}{1988}]{owocki1988}
{Owocki}, S.~P., {Castor}, J.~I., {Rybicki}, G.~B. 1988, \apj, 335, 914

\bibitem[\protect\astroncite{{Owocki} \& {Cohen}}{2006}]{owocki2006}
{Owocki}, S.~P. \& {Cohen}, D. 2006, \apj, astroph/0602o54

\bibitem[\protect\astroncite{{Parmar} et~al.}{1984}]{parmar1984}
{Parmar}, A.~N., {Culhane}, J.~L., {Rapley}, C.~G., {Wolfson}, C.~J., {Acton}, L.~W.,
{Phillips}, K.~J.~H., {Dennis}, B.~R. 1984, \apj, 279, 866

\bibitem[\protect\astroncite{{Pittard} et~al.}{2000}]{pittard2000}
{Pittard}, J.~M., {Stevens}, I.~R., {Corcoran}, M.~F., 
{Gayley}, K.~G., {Marchenko}, S.~V., {Rauw}, G. 2000, \mnras, 319, 137

\bibitem[\protect\astroncite{{Pollock}}{1987}]{pollock1987}
{Pollock}, A.~M.~T. 1987, \apj, 320, 283

\bibitem[\protect\astroncite{{Pollock} et~al.}{2005}]{pollock2005}
{Pollock}, A.~M.~T., {Corcoran}, M.~F., {Stevens}, I.~R. \& {Williams}, P.~M. 2005, \apj, 629, 482

\bibitem[\protect\astroncite{{Porquet} et~al.}{2001}]{porquet2001}
{Porquet}, D., {Mewe}, R., {Dubau}, J., {Raassen}, A.~J.~J., {Kaastra}, J.~S. 2001, \aap, 376, 1113

\bibitem[\protect\astroncite{{Preibisch} et~al.}{1999}]{preibisch1999}
{Preibisch}, T., {Balega}, Y., {Hoffmann}, K.-H., {Weigelt}, G., {Zinnecker}, H. 1999,
New A., 4, 531

\bibitem[\protect\astroncite{{Prilutskii} \& {Usov}}{1976}]{prilutskii1976}
{Prilutskii}, O.~F. \& {Usov}, V.~V. 1976, \azh, 53, 6

\bibitem[\protect\astroncite{{Rauw} et~al.}{2002}]{rauw2002}
{Rauw}, G., {Vreux}, J.-M., {Stevens}, I. R., {Gosset}, E., {Sana}, H., {Jamar}, C., 
{Mason}, K. O. 2002, \aap, 388, 552

\bibitem[\protect\astroncite{{Sako} et~al.}{1998}]{sako1999}
{Sako}, M., {Liedahl}, D.~A., {Kahn}, S.~M., {Paerels}, F. 1999, \apj, 525, 921

\bibitem[\protect\astroncite{{Sana} et~al.}{2004}]{sana2004}
{Sana}, H., {Stevens}, I. R., {Gosset}, E., {Rauw}, G., {Vreux}, J.-M. 2004, \mnras, 350, 809

\bibitem[\protect\astroncite{{Schulz} et~al.}{2000}]{schulz2000}
{Schulz}, N.~S., {Canizares}, C., {Huenemoerder}, D., \& {Lee}, J. 2000,
  \apj, 545, L135

\bibitem[\protect\astroncite{{Schulz} et~al.}{2001}]{schulz2001}
{Schulz}, N.~S., {Canizares}, C., {Huenemoerder}, D., {Kastner}, J.~H., {Taylor}, S.~C.,
\& {Bergstrom}, E.~J. 2001, \apj, 549m 441

\bibitem[\protect\astroncite{{Schulz} et~al.}{2002}]{schulz2002}
{Schulz}, N.~S., {Canizares}, C.~R., {Lee}, J.~C., {Sako}, M. 2002, \apj, 564, 21

\bibitem[\protect\astroncite{{Schulz} et~al.}{2003}]{schulz2003}
{Schulz}, N.~S., {Canizares}, C., {Huenemoerder}, D., \& {Tibbets}, K. 2003,
  \apj, 595, 365

\bibitem[\protect\astroncite{{Schulz}}{2003}]{schulz2003a}
{Schulz}, N.~S. 2003, R.Mx.A.C., 15, 220

\bibitem[\protect\astroncite{{Seward} et~al.}{1979}]{seward1979}
{Seward}, F.~D., {Forman}, W.~R., {Giaconni}, R., {Griffith}, R.~E., {Harnden}, F.~R.,
{Jones}, C., {Pye}, J.~P. (1979), \apj, 234, L55

\bibitem[\protect\astroncite{{Shu} et~al.}{1997}]{shu1997}
{Shu}, F.~H., {Shang}, H., {Glassgold}, A.~E., \& Lee, T. 1997, Science, 277, 1475

\bibitem[\protect\astroncite{{Smith} et~al.}{2001}]{smith2001}
{Smith}, R.~K, {Brickhouse} N.~S., {Liedahl}, D.~A., \& {Raymond} J.~C. 2001, \apj, 556, 91

\bibitem[\protect\astroncite{{Stelzer} et~al.}{2005}]{stelzer2005}
{Stelzer}, B., {Flaccomio}, E., {Montmerle}, T., {Micela}, G., {Sciortino}, S., {Favata}, F., 
{Preibisch}, T., \& {Feigelson}, E.~D. 2005, \apj Suppl., 160, 557

\bibitem[\protect\astroncite{{Stevens} et~al.}{1992}]{stevens1992}
{Stevens}, I. R., {Blondin}, J. M., \& {Pollock} A. M. T. 1992, \apj, 386, 265

\bibitem[\protect\astroncite{{Testa} et~al.}{2004}]{testa2004}
{Testa}, P., {Drake}, J.~J., \& {Peres}, G. 2004, \apj, 617, 508

\bibitem[\protect\astroncite{{Townsley} et~al.}{2003}]{townsley2003}
{Townsley}, L.~K., {Feigelson}, E.~D., {Montmerle}, T., {Broos}, P.~S., {Chu}, Y.-H.,
{Garmire}, G.~P. 2003, \apj, 593, 874

\bibitem[\protect\astroncite{{Tsujimoto} et~al.}{2005}]{tsujimoto2005}
{Tsujimoto}, M., {Feigelson}, E.~D., {Grosso}, N., {Micela}, G., {Tsuboi}, Y., {Favata}, F.,
{Shang}, H., {Kastner}, J.~H. 2005, \apj, 160, 503

\bibitem[\protect\astroncite{{ud-Doula} \& {Owocki}}{2002}]{uddoula2002}
{ud-Doula}, A., {Owocki}, S.~P. 2002, \apj, 576, 413

\bibitem[\protect\astroncite{{Waldron} et~al.}{2004}]{waldron2004}
{Waldron}, W.~L., {Cassinelli}, J.~P., {Miller}, N.~A., {MacFarlane}, J.~J., {Reiter}, J.~C.
2004, \apj, 616,  542

\bibitem[\protect\astroncite{{Waldron} \& {Cassinelli}}{2001}]{waldron2001}
{Waldron}, W.~L., {Cassinelli}, J.~P. 2001, ApJ, 548, 45

\bibitem[\protect\astroncite{{Wojdowski} \& {Schulz}}{2005}]{wojdowski2005}
{Wojdowski}, P.~S. \& {Schulz}, N.~S. 2005, \apj, 627, 953

\end{thebibliography}
\end{document}